\documentclass[sigconf, nonacm=true]{acmart}

\makeatletter
\def\@ACM@checkaffil{
    \if@ACM@instpresent\else
    \ClassWarningNoLine{\@classname}{No institution present for an affiliation}%
    \fi
    \if@ACM@citypresent\else
    \ClassWarningNoLine{\@classname}{No city present for an affiliation}%
    \fi
    \if@ACM@countrypresent\else
        \ClassWarningNoLine{\@classname}{No country present for an affiliation}%
    \fi
}
\makeatother

\usepackage{multirow}
\usepackage{amsmath, amsfonts}
\usepackage{bbold}
\usepackage{makecell}
\usepackage{enumitem}

\usepackage{pifont}
\newcommand{\cmark}{\ding{51}}%
\newcommand{\xmark}{\ding{55}}

\newcommand{\hquad}{\hspace{0.5em}} 

\AtBeginDocument{%
  }

\begin{document}

\title{\textit{Ab-initio} Quantum Transport with the \textit{GW} Approximation, 42,240 Atoms, and Sustained Exascale Performance}

\author{Nicolas Vetsch$^{*}$, Alexander Maeder$^{*}$, Vincent Maillou$^{*}$, Anders Winka$^{*}$, Jiang Cao$^{*}$, Grzegorz Kwasniewski$^{\dagger}$, Leonard Deuschle$^{*}$, Torsten Hoefler$^{\dagger}$, Alexandros N. Ziogas$^{*}$, and Mathieu Luisier$^{*}$} 
\affiliation{
$^*$Integrated Systems Laboratory, ETH Zurich, Switzerland\\
$^{\dagger}$Scalable Parallel Computing Laboratory, ETH Zurich, Switzerland 
}

\renewcommand{\shortauthors}{Vetsch et al.}

\begin{abstract}
Designing nanoscale electronic devices such as the currently manufactured nanoribbon field-effect transistors (NRFETs) requires advanced modeling tools capturing all relevant quantum mechanical effects. State-of-the-art approaches combine the non-equilibrium Green's function (NEGF) formalism and density functional theory (DFT). However, as device dimensions do not exceed a few nanometers anymore, electrons are confined in ultra-small volumes, giving rise to strong electron-electron interactions. To account for these critical effects, DFT+NEGF solvers should be extended with the \textit{GW} approximation, which massively increases their computational intensity. Here, we present the first implementation of the NEGF+\textit{GW} scheme capable of handling NRFET geometries with dimensions comparable to experiments. This package, called \textit{QuaTrEx}, makes use of a novel spatial domain decomposition scheme, can treat devices made of up to 84,480 atoms, scales very well on the Alps and Frontier supercomputers ($>$80\% weak scaling efficiency), and sustains an exascale FP64 performance on 42,240 atoms (1.15 Eflop/s).
\end{abstract}

\maketitle

\section{Justification for ACM Gordon Bell Prize}
We report \textit{ab-initio} transistor simulations of unprecedented scale (up to 84,480 atoms) including electron-electron interactions within the self-consistent \textit{GW} approximation. Key achievements are simulations of 42,240 atoms on 37,600 GPUs with (i) ultra-short iteration time ($\sim$42 seconds per iteration), (ii) excellent parallel efficiency (82$\%$ in weak scaling), and (iii) high computational performance (1.15 Eflop/s --- 85\% of Rmax --- 56\% of Rpeak).

\section{Performance Attributes}

\begin{table}[h]
	\small
\begin{tabular}{lp{4cm}} 
\toprule
Performance attributes & Our submission \\
\midrule
Category of achievement & Peak performance, time-to-solution\\
Type of method used & Non-linear system of equations\\
Results reported on basis of & Full application except I/O\\
Precision reported & Double precision\\
System scale & Full-scale\\
Measurements & Timers, FLOP count\\
\bottomrule
\end{tabular}
	\vspace{-1em}
\end{table}

\section{Overview of the Problem}

\begin{figure}[t]
    \centering
    \includegraphics[width=\columnwidth]{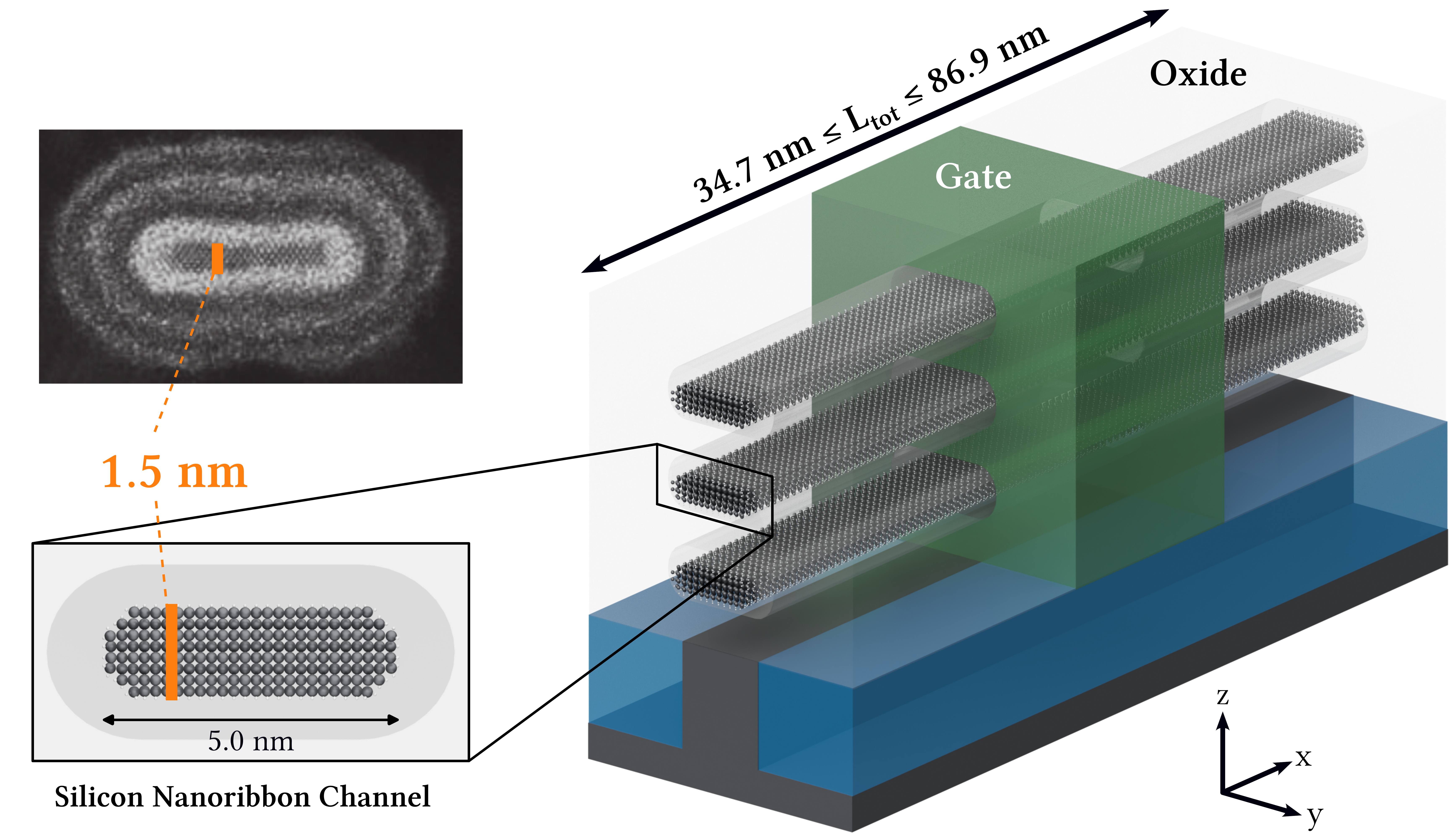}
    \caption{Schematic view of a silicon nanoribbon field-effect transistor where three nanoribbon layers are stacked on top of each other. The central ribbon is simulated in this work. Its cross section is shown on the left, highlighting its shape and size resemblance with a device structure that was recently fabricated by Intel. Reprinted with permission from \cite{intelnr}.}
    \label{fig:ribbonfet}
\end{figure}

Density functional theory (DFT) \cite{kohnsham} is the method of choice to determine the quantum mechanical properties of metals, oxides, and semiconductors, which build the core of all (opto-)electronic devices, e.g., transistors, light-emitting diodes, or memory cells. Physical quantities such as their band gap, effective mass, defect forming energy, structure stability, or absorption coefficient can be readily obtained with DFT tools, for example, VASP~\cite{vasp}, GPAW~\cite{GPAW}, Quantum Espresso~\cite{qe}, Abinit~\cite{abinit}, CP2K~\cite{cp2k}, or Siesta~\cite{siesta}. Although DFT is a powerful \textit{ab-initio} framework with the capability of predicting several material properties, it remains a ground-state theory and, as such, does not describe excited electronic states very accurately \cite{Perdew2009}. In particular, it underestimates the band gap of most common semiconductors (Si, Ge, or GaAs) and oxides (SiO$_2$, HfO$_2$, or Al$_2$O$_3$) by a factor up to 2, band alignments between different compounds are not correctly reproduced, while important effects like Auger processes or excitons are missing \cite{Onida2002}.

Modeling approaches going beyond DFT have, therefore, been developed to overcome these limitations. The \textit{GW} approximation is one of them. It relies on Hedin's equations, which establish a relationship between electrons, whose behavior is described by a variable called $G$, and the (screened) Coulomb interactions they induce (attraction/repulsion forces) labeled as $W$~\cite{Hedin1965}. The product of $G$ and $W$ does not only give the \textit{GW} method its name, but also forms a self-energy $\Sigma$ that corrects DFT results, in particular band gaps~\cite{Hybertsen1986}, thus providing optical absorption spectra \cite{Qiu2013} or defect levels \cite{Chen2017} in excellent agreement with experimental data.

\begin{table*}[!t]
	\caption{Summary of the state-of-the-art \textit{GW} performance. The tool name, type of \textit{GW} approximation, inclusion of transport properties, basis set into which the Hamiltonian matrices are expanded, computational scalability, number of atoms achieved on a specific hardware, and simulation time are reported. All symbols are defined in the text and in Table \ref{tab:structure_param}.
    }
    \small
    \vspace{-1em}
    \def\arraystretch{1.1}
	\begin{tabular}{l c c c c c c c c}
    \multirow{2}{*}{\textbf{Tool}}&\multirow{2}{*}{\textbf{Type}}&\multirow{2}{*}{\textbf{Transport}}&\multirow{2}{*}{\textbf{Basis}}&\multirow{2}{*}{\textbf{Scalability}}&\multirow{2}{*}{\textbf{No. Atoms}}&\textbf{Performance}&\multirow{2}{*}{\textbf{Hardware}}&\textbf{Time$^*$}\\
    &&&&&&\textbf{[Pflop/s]}&& \textbf{[s]}\\
    \hline
    VASP \cite{vaspgw}& \textit{G$_0$W$_0$} & No & PAW & $\mathcal{O}(N_A^3)$ & 54 & -- & 64 CPUs & 720\\
    CP2K \cite{cp2kgw}& \textit{G$_0$W$_0$} & No & GTO & $\mathcal{O}(N_A^{2.13})$ & 1,734 & -- & 14,400 CPUs &18,000\\
    BerkeleyGW \cite{berkeleygw}& \textit{G$_0$W$_0$} & No & PW & $\mathcal{O}(N_A^{4})$ & 2,742 & 105.9 & 27,648 GPUs & $>$1,000\\
    WEST \cite{westgw} & \textit{G$_0$W$_0$} & No & PW & $\mathcal{O}(N_A^{4})$& 1,728 & 36-59 & 25,960 GPUs & $>$1,500\\
    NanoGW \cite{nanogw} & \textit{G$_0$W$_0$} & No & RSG & $\mathcal{O}(N_A^{4})$ & 2,551 & -- &1,280 CPUs + 80 GPUs & 33,120\\
    PWDFT \cite{gwsc24}& \textit{G$_0$W$_0$} & No & PW & $\mathcal{O}(N_A^3)$ & 13,824 & -- & 449,280 cores & 285\\
    FlapwMBPT \cite{scgw}& sc\textit{GW}$^{\dagger}$ & No & RSG & $\mathcal{O}(N_A)$ to $\mathcal{O}(N_A^2)$& 72 & -- & 288 CPUs & 18,000\\
    QuaTrEx$_{24}$ \cite{deuschle2024_sc}& sc\textit{GW}$^{\dagger}$ & NEGF & MLWF & $\mathcal{O}(N_EN_BN_{BS}^3)$ & 10,560 & 69.3 & 7,200 GPUs & 31\\
    \hline
    \multirow{2}{*}{\textbf{This work}} & \multirow{2}{*}{\textbf{sc\textit{GW}$^{\dagger}$}} & \multirow{2}{*}{\textbf{NEGF}} & \multirow{2}{*}{\textbf{MLWF}} & \multirow{2}{*}{\textbf{$\mathcal{O}(N_EN_BN_{BS}^3)$}} & \textbf{46,464} & \textbf{342.64} & \textbf{9,400 GPUs} & \textbf{25} \\
    & & & & & \textbf{42,240} & \textbf{1,146.04} & \textbf{37,600 GPUs} & \textbf{42} \\
    \hline
    \end{tabular}\\
    {\footnotesize $^*$: Estimates based on data provided, except for QuaTrEx$_{24}$, PWDFT, and this work (measurements). $\dagger$: Scalability and time are given per iteration.}\vspace{-1em}
	\label{tab:competitors}
\end{table*}

While the extension of DFT with \textit{GW} enables studying the properties of materials with high accuracy, it fails at capturing non-equilibrium phenomena such as the voltage-driven electronic current flowing through atomic systems, i.e., their ``current vs. voltage'' characteristics. Combining DFT and the \textit{GW} approximation with the non-equilibrium Green's function (NEGF) formalism addresses this issue and, therefore, allows for the exploration of the quantum transport properties of devices at the \textit{ab-initio} level, with an atomistic resolution, and in the presence of electron-electron interactions~\cite{thygesen2007}. The latter might play a critical role in future transistors with ultra-short gate lengths. Contrary to expectations, these devices will probably not deliver electronic currents that are closer to their theoretical (ballistic) limit than previous generations~\cite{Fischetti2001}. Although electrons (or holes) have a lower probability to interact with crystal vibrations (phonons), rough surfaces, or defects when gate lengths do not exceed 10-15 nm, the densely populated source and drain electrodes of these transistors become so close to each other that the carriers located there start to strongly interact. Additional scattering that does not exist in longer-channel devices is thus induced. It might severely limit the drive current of, for example, nanoribbon field-effect transistors (NRFET), as illustrated in Fig.~\ref{fig:ribbonfet}, where the carrier population is highly confined. Hence, the inclusion of electron-electron interactions through the $GW$ approximation is indispensable to predict the performance of not-yet-fabricated devices and to guide the design of next-generation transistors with sufficiently high output currents.

The major bottleneck of NEGF+\textit{GW} approaches is their computational intensity.
When DFT+NEGF \textit{ab-initio} quantum transport solvers were pioneered in the early 2000s, without \textit{GW} corrections, they were limited to tens of atoms \cite{stokbro}. Algorithmic and hardware developments have pushed back this limit to 10,000s of atoms, still in the ballistic limit of transport (no electron interactions with other particles)~\cite{calderara}. More recently, by leveraging data-centric programming, Joule heating could be introduced into a DFT+NEGF tool to treat realistic transistor geometries on GPUs~\cite{ziogas}. However, self-consistently solving for the electron and phonon (crystal vibration) populations increases the workload by almost two orders of magnitude as compared to ballistic transport. Climbing up the ladder of physical and computational complexity, the inclusion of electron-electron interactions within the \textit{GW} approximation is the next step. This relevant feature was demonstrated in 2024 for nanowire test structures made of up to 10,000 atoms~\cite{deuschle2024_sc} but not for devices comparable to those produced in semiconductor fabs.

Here, we present an optimized implementation of the NEGF+\textit{GW} scheme where each computational kernel is carefully optimized to take advantage of the Alps~\cite{hoefler_alps} and Frontier~\cite{Atchley2023_frontier} supercomputers. Key improvements have been made in the following areas: 
\begin{itemize}[leftmargin=*]
\item Spatial domain decomposition: Development and integration of a distributed linear solver to compute selected entries of the Green's function and screened Coulomb interaction matrices, allowing for the first simulation of devices with up to 84,480 atoms, in the presence of electron-electron interactions;
\item Memory management: Exploitation of the symmetry of all involved physical quantities to reduce the memory footprint, and introduction of a memoization technique to boost specific, otherwise time-consuming calculations;
\item Sustained performance: Python-orchestrated code running on Alps and Frontier up to 9,400 and 37,600 GPUs, reaching a FP64 performance of 343 Pflop/s and 1.146 Eflop/s, respectively;
\item Time-to-solution: Investigation of realistic NRFETs with dimensions comparable to experimental devices and generating 16$\times$ the simulation workload of the state of the art, with only a 35\% increase in computational time (42.1~sec per iteration vs. 31.3~sec).
\end{itemize}

\subsection{The \textit{GW} Approximation}
Although Hedin's equations have been known for 60 years, successful implementations of the \textit{GW} approximation have not been realized before 1986 and the seminal work by Hybertsen and Louie \cite{Hybertsen1986}. The general approach consists of first computing a Green's function $G$ from DFT wavefunctions $\psi$, then a screened Coulomb interaction $W$, and finally a self-energy $\Sigma=GW$ that modifies the original DFT results. Although the procedure is self-consistent ($G$ depends on the $\psi$, which are in turn altered by $\Sigma$), it is usually restricted to a one-shot calculation due to its high computational ($\mathcal{O}(N_A^4)$) and memory ($\mathcal{O}(N_A^3)$) requirements, $N_A$ being the total number of atoms in the system of interest. This is known as the $G_0W_0$ approximation, where $G$ and $W$ are only evaluated once.

To treat large structures, attempts have been made to improve the scalability of the $G_0W_0$ method. The developers of VASP~\cite{vasp}, a package relying on a plane-wave (PW) basis, managed to decrease the computational complexity from $\mathcal{O}(N_A^4)$
to $\mathcal{O}(N_A^3)$ \cite{vaspgw}, while the CP2K team took advantage of the high localization of their Gaussian-type orbitals (GTO) to reach $\mathcal{O}(N_A^{2.13})$ and simulate 1,734 atoms \cite{cp2kgw}. Alternatively, optimizations in the implementation of the $G_0W_0$ approximation have been reported, without changing its scalability. BerkeleyGW (2,742 silicon atoms in a PW basis) \cite{berkeleygw}, WEST (1,728 silicon atoms with PW) \cite{westgw}, and NanoGW (silicon quantum dot comprising 2,551 atoms in a real-space grid (RSG)) \cite{nanogw} are excellent examples of recent progress in this area. Finally, combinations of algorithmic innovations and software developments have led to the calculation of systems with up to 13,824 atoms using the PWDFT tool \cite{gwsc24}.

Going beyond the $G_0W_0$ approximation in terms of physical accuracy requires solving $G$ and $W$ self-consistently, which can take hundreds of iterations to converge. The resulting self-consistent $GW$ (sc$GW$) schemes scale with $\mathcal{O}(N_{iter}N_A^{4})$, where $N_{iter}$ is the number of iterations needed to attain convergence. Despite significant scalability improvements, e.g., $\mathcal{O}(N_A)$ to $\mathcal{O}(N_A^2)$
for \mbox{FlapwMBPT}~\cite{scgw}, the capabilities of sc$GW$ solvers rarely exceed tens of atoms. At the next level of complexity, the non-equilibrium Green's function formalism is combined with the self-consistent $GW$ approximation to capture correlation effects in nano-devices driven out of equilibrium by external voltages, as the NRFET in Fig.~\ref{fig:ribbonfet}. In 2007, it was shown that this can be done in small molecules \cite{thygesen2007}. In 2024, we investigated a silicon nanowire composed of 10,560 atoms with NEGF+sc\textit{GW} and the first prototype of our quantum transport tool QuaTrEx \cite{deuschle2024_sc}. Here, we refer to this version of the code from 2024 as QuaTrEx$_{24}$ and to the current one simply as QuaTrEx. Major achievements in the $GW$ approximation are summarized in Table \ref{tab:competitors}.

\subsection{The NEGF+sc\textit{GW} Scheme}\label{sec:quantum-transport}
Contrary to equilibrium $G_0W_0$ or sc$GW$ approaches, NEGF+sc\textit{GW} relies on two particle descriptors ($G$ / $W$) and two interaction terms ($\Sigma$ / $P$) of four different types; retarded~($R$), advanced~($A$), lesser~($<$), and greater~($>$). All of them are expressed in a basis of localized atomic orbitals. Their governing matrix equations can be written in the following compact form \cite{deuschle2024_arXiv} 
\begin{equation}
\left[\mathbf{M}(E)-\mathbf{B}^{R}(E)\right]\mathbf{X}^{\lessgtr}(E)\left[\mathbf{M}(E)-\mathbf{B}^{R}(E)\right]^{\dagger} = \mathbf{B}^{\lessgtr}(E),
\label{eq:system-solve}
\end{equation}
\begin{equation}
\mathbf{B}(E) = \mathbf{B}_{OBC}(E) + \mathbf{B}_{scatt}(E),
\label{eq:B}
\end{equation}
\begin{equation}
\mathbf{B}_{scatt}(E) \propto \int dE' \mathbf{X}_{1}(E-E') \mathbf{X}_{2}(E').
\label{eq:interaction}
\end{equation}
The $\mathbf{X}_{(i)}$ matrices refer to the particle descriptors $\mathbf{G}$ or $\mathbf{W}$, while $\mathbf{B}$ corresponds to the interaction terms $\mathbf{P}$ or $\mathbf{\Sigma}$. All these quantities are functions of energy $E$, and the above set of equations typically has to be solved for $10,000\leq N_E\leq 100,000$ energy points. Furthermore, since $\mathbf{G}$, $\mathbf{P}$, $\mathbf{W}$, and $\mathbf{\Sigma}$ are interdependent, their equations must be repeatedly evaluated until reaching self-consistency. This iterative approach is known as the self-consistent Born approximation (SCBA). Up to $N_{iter}$=500 iterations are needed to achieve convergence with it. It should be noted that $\mathbf{B}$ contains two parts, one labeled $\mathbf{B}_{OBC}$ that accounts for the so-called open boundary conditions (OBCs) and connects the device with its contacts, and one denoted $\mathbf{B}_{scatt}$ to model electron-electron scattering. 

\begin{table}[h]
\caption{Definition of the physical quantities in Eqs.~(\ref{eq:system-solve}) to (\ref{eq:interaction}).}
        \vspace{-1em}
\small
\def\arraystretch{1.3}
    \centering
    \begin{tabular}{c|c|c}
        Quantity & Electrons & Screened Coulomb\\
        \hline
        $\mathbf{M}(E)$ & $ E\mathbf{S}_{\mathbf{DFT}} - \mathbf{H}_{\mathrm{DFT}}$ & $\mathbf{I}$\\
        $\mathbf{B}^{R}_{scatt}(E)$  & $\mathbf{\Sigma}^{R}(E)$ & $\mathbf{V}\mathbf{P}^{R}(E)$ \\
        $\mathbf{B}^{\lessgtr}_{scatt}(E)$  & $\mathbf{\Sigma}^{\lessgtr}(E)$ & $\mathbf{V}\mathbf{P}^{\lessgtr}(E)\mathbf{V}^{\dagger}$ \\
        \hline
        $\widetilde{\mathbf{M}}(E)$ & \multicolumn{2}{c}{$\mathbf{M}(E) - \mathbf{B}^{R}_{scatt}(E) - \mathbf{B}^{R}_{OBC}(E)$}\\
        \hline
       \end{tabular}
    \label{tab:quantities}
\end{table}

The link between Eqs.~(\ref{eq:system-solve}) to (\ref{eq:interaction}) and all underlying physical quantities is provided in Table \ref{tab:quantities}. As NEGF+sc$GW$ is an extension to DFT, it takes inputs from it, namely the Hamiltonian ($\mathbf{H}_{\mathbf{DFT}}$), overlap ($\mathbf{S}_{\mathbf{DFT}}$), and bare Coulomb ($\mathbf{V}$) matrices.

\section{Current State of the Art}
Here, we describe the main computational tasks of an NEGF+sc\textit{GW} solver and discuss state-of-the-art algorithms to accomplish them.

As already mentioned, two mutually interacting subsystems are coupled in NEGF+sc\textit{GW}, the electron population and the screened Coulomb interaction. For each of them, three main tasks are sequentially executed after gathering input data from DFT. First, to inject particles into the simulation domain and collect them when they exit, suitable OBCs must be constructed. The second step consists of assembling the system's governing matrix equation and computing its solution for all energies that the particle descriptor $\mathbf{G}(E)$ or $\mathbf{W}(E)$ can occupy. In a last step, the interactions between them, i.e., $\mathbf{P}(E)$ and $\mathbf{\Sigma}(E)$, are determined via energy convolution. 

Before describing these three tasks and the associated algorithms in detail, we elaborate on the structure of the data we are dealing with in the NEGF+sc\textit{GW} method.

\begin{table*}[h]
\caption{Physical dimensions of the nano-device structures considered in this work (NW: nanowire, NR: nanoribbon). The parameters characterizing their numerical representation within the NEGF formalism are listed in the bottom part of the table.}
\label{tab:structure_param}
\vspace{-1em}
\small
\centering
\begin{tabular}{ l l l c c c c c c}
    \hline
    Variable & Description & Formulation & NW-1 & NW-2 & NR-16 & NR-24 & NR-40 & NR-$N_B$ \\ 
    \hline
    & & & \raisebox{-.5\height}{\includegraphics[scale=0.15]{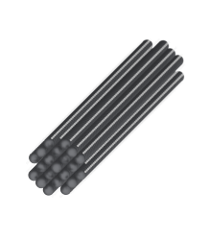}} & \raisebox{-.5\height}{\includegraphics[scale=0.15]{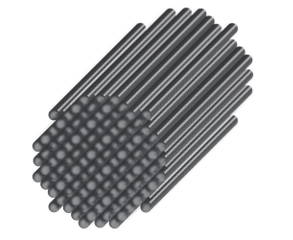}} & \multicolumn{4}{c}{\rule[1.5pt]{0.75cm}{0.4pt}~\raisebox{-.5\height}{\includegraphics[scale=0.15]{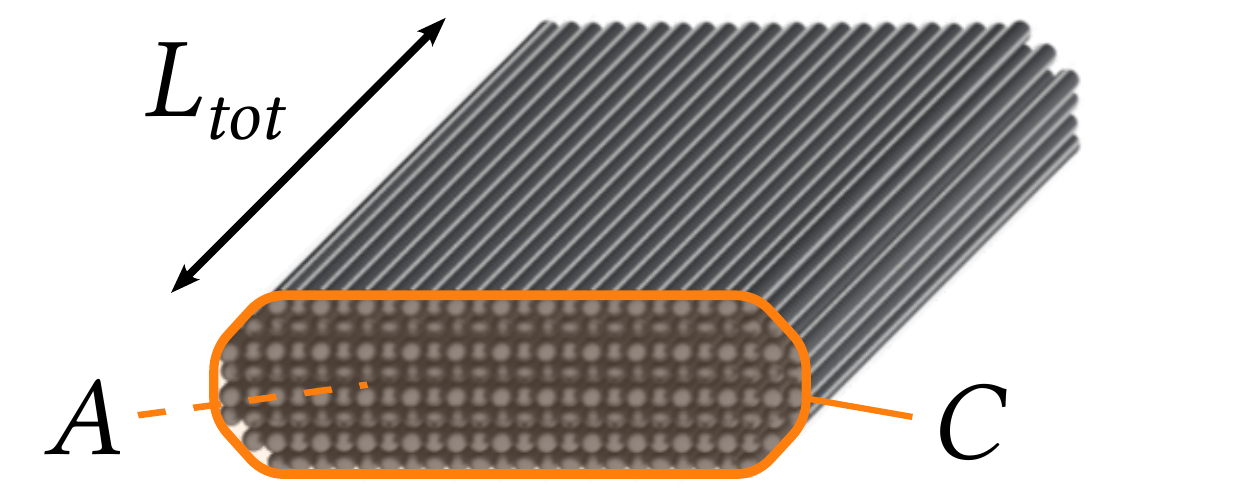}}~\rule[1.5pt]{0.75cm}{0.4pt}}\\
    \hline
    $L_{tot}$ & \footnotesize Device length (nm) \small & & 39.1 & 34.7 & 34.7 & 52.1 & 86.9 & 2.172$\times N_B$ \\
    $A$ & \footnotesize Device cross section (nm$^2$) \small & & 0.8 & 4.3 & \multicolumn{4}{c}{\rule[1.5pt]{2.2cm}{0.4pt}~7.5~\rule[1.5pt]{2.2cm}{0.4pt}} \\
    $C$ & \footnotesize Device circumference (nm) \small & & 3.1 & 6.9 & \multicolumn{4}{c}{\rule[1.5pt]{2.15cm}{0.4pt}~13.0~\rule[1.5pt]{2.15cm}{0.4pt}} \\
    $r_{cut}$ & \footnotesize Interaction cutoff distance (\AA) \small & & 10.95 & 7.15 & \multicolumn{4}{c}{\rule[1.5pt]{2.15cm}{0.4pt}~7.5~\rule[1.5pt]{2.15cm}{0.4pt}} \\
    \hline 
    $N_{A}$ & \footnotesize Number of atoms \small & $\mathcal{O}\left(\left(A+C\right)L_{tot}\right)$ & 2,952 & 10,560 & 16,896 & 25,344 & 42,240 & 1,056$\times N_B$  \\
    $N_{AO}$ & \footnotesize Number of atomic orbitals \small & $\mathcal{O}\left(\left(A+C\right)L_{tot}\right)$ & 7,488 & 32,256 & 54,528 & 81,792 & 136,320 & 3,408$\times N_B$  \\
    $\widetilde{N}_{BS}$ & \footnotesize Primitive unit cell size \small & $\mathcal{O}\left(A\right)$ & 104 & 504 &  \multicolumn{4}{c}{\rule[1.5pt]{2.15cm}{0.4pt}~852~\rule[1.5pt]{2.15cm}{0.4pt}}  \\
    $N_U$ & \makecell[l]{\footnotesize Number of primitive unit cells \\ \footnotesize per transport cell $\left(\mathbf{G}/\mathbf{W}\right)$} \small & MLWF-dependent & 4/8 & \multicolumn{5}{c}{\rule[1.5pt]{3.05cm}{0.4pt}~4~\rule[1.5pt]{3.05cm}{0.4pt}} \\
    $N_{BS}$ & \footnotesize Transport cell size for \textbf{G}/\textbf{W} \small & $\widetilde{N}_{BS}N_U$& 416/832 & 2,016 & \multicolumn{4}{c}{\rule[1.5pt]{2.11cm}{0.4pt}~3,408~\rule[1.5pt]{2.1cm}{0.4pt}} \\
    $N_{B}$ & \footnotesize Number of transport cells $\left(\mathbf{G}/\mathbf{W}\right)$ \small & $N_{AO}/N_{BS}$ & 18/9 & 16 & 16 & 24 & 40 & $N_B$  \\
    $H_{NNZ}$ & \footnotesize Number of non-zeros$^*$ in \textbf{H} \small & $\mathcal{O}\left(N_U\widetilde{N}_{BS}N_{AO}\right)$ & $0.5\times10^7$ & $14.1\times10^7$ & $40.4\times10^7$ & $61.3\times10^7$ & $103.1\times10^7$ & $\approx2.6N_B\times10^7$ \\
    $G_{NNZ}$ & \makecell[l]{\footnotesize Number of non-zeros$^*$ in the \\ \footnotesize \textbf{G}, \textbf{P}, \textbf{W}, and $\mathbf{\Sigma}$ matrices \small}& $\mathcal{O}\left(r_{cut}\widetilde{N}_{BS}N_{AO}\right)$ &  $0.3\times10^7$ & $4.3\times10^7$ & $12.6\times10^7$ & $19.0\times10^7$ & $31.8\times10^7$ & $\approx0.8N_B\times10^7$ \\
 \hline
\end{tabular}\\
{\footnotesize $^*$Number of non-zeros are given when no symmetry applies.}\vspace{-1em}
\end{table*}

\begin{figure}[h]
    \centering
    \includegraphics[width=0.9\columnwidth]{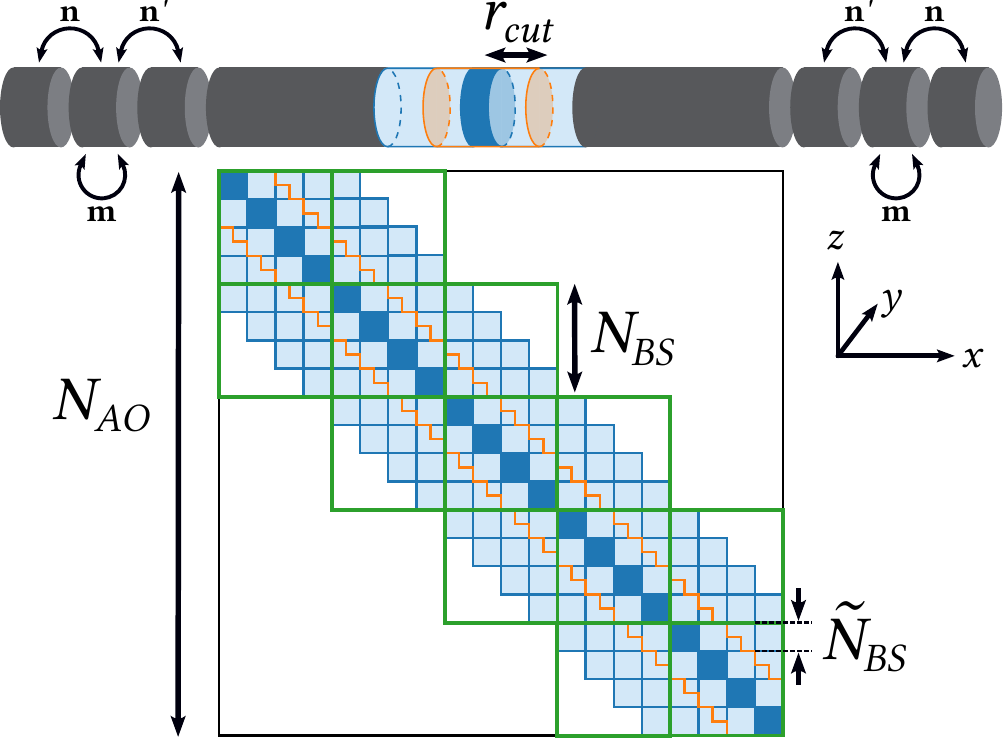}
    \caption{Schematic mapping of a general nanowire structure (top) onto a block-banded matrix of size $N_{AO}\times N_{AO}$. The sparsity pattern can be tiled as a block-tridiagonal matrix with $N_B$ diagonal blocks of size $N_{BS}$ (green boxes). Each of them can be further decomposed into smaller blocks of size $\widetilde{N}_{BS}$ corresponding to the primitive unit cell of the nanostructure. The effect of introducing a cut-off radius $r_{cut}$ for the inter-atomic interactions is shown in orange, both in the matrix and device structure. The periodic contact layers are marked on the left and right of the nanowire.}

    \label{fig:device_to_bt_matrix}
\end{figure}

\subsection{Matrix Representations of Nano-devices}\label{sec:data-structure}
\label{sec:datastructures}
When simulating nano-devices out of equilibrium, the structure's periodicity is broken along at least one dimension by the externally applied voltage. This defines the transport axis, e.g., $x$, along which carriers (electrons and/or holes) propagate. As a consequence, spatially extended basis functions such as the plane-waves commonly used in DFT and $GW$ (see Table~\ref{tab:competitors}) are not suitable for NEGF. A sufficiently localized basis set is needed to construct the $\mathbf{H}_{\mathbf{DFT}}$, $\mathbf{S}_{\mathbf{DFT}}$, and $\mathbf{V}$ matrices. CP2K \cite{cp2k} or Siesta \cite{siesta} natively provide such localized functions. Alternatively, maximally localized Wannier functions (MLWF) \cite{mlwf} can be created from a PW basis with Wannier90 \cite{Wannier90}.

In this work, we consider eight nano-devices whose structure models a silicon channel passivated with hydrogen atoms (see Table \ref{tab:structure_param}). Transport occurs along the $x$ axis, while $y$ and $z$ are directions of confinement. We take advantage of MLWF to build their $\mathbf{H}_{\mathbf{DFT}}$ and $\mathbf{V}$ matrices ($\mathbf{S}_{\mathbf{DFT}}=\mathbf{I}$ because of the orthogonality of our MLWF basis). Using the PW DFT code VASP~\cite{vasp}, we first compute the electronic structure of a primitive unit cell (PUC) that can be repeated along the transport axis to obtain the desired device length $L_{tot}$. The results are converted into MLWFs (4 per Si, 1 per H).

The outcome is a square Hamiltonian matrix $\mathbf{h}_{ii}$ ($\mathbf{h}_{ij}$) of size $\widetilde{N}_{BS}\times\widetilde{N}_{BS}$ that represents the coupling between all MLWFs located within the original PUC with themselves (with MLWFs situated in neighboring cells), as illustrated in Fig.~\ref{fig:device_to_bt_matrix}. In this example, it can be seen that one PUC (dark blue) is connected to four others along the $+x$ axis ($\mathbf{h}_{ii+1}$ to $\mathbf{h}_{ii+4}$, light blue). Since the magnitude of the interactions between two MLWF situated further apart rapidly decreases, the $\mathbf{h}_{ij}$ blocks with $j>4$ can be safely ignored.

Because (1) $\mathbf{H}_{\mathbf{DFT}}$ is per definition Hermitian and (2) our test structures are made of the repetition of a single PUC along $x$, the knowledge of the five aforementioned $\mathbf{h}_{ij}$ blocks is sufficient to construct the Hamiltonian matrix of any device with length $L_{tot}$ and total number of atoms $N_A$. The size of $\mathbf{H}_{\mathbf{DFT}}$ is $N_{AO}\times N_{AO}$, where $N_{AO}$ denotes the number of MLWFs in the system. We finally arrive at a block-banded (BB) sparsity pattern for the full Hamiltonian matrix $\mathbf{H}_{\mathbf{DFT}}$, as depicted in Fig.~\ref{fig:device_to_bt_matrix}.

The bare Coulomb matrix $\mathbf{V}$ can be directly computed in the MLWF basis \cite{thygesen2007}. It is full, as electron-electron interactions typically extend over long ranges. Nevertheless, the magnitude of the $V_{ij}$ entries between two basis functions $i$ and $j$ decays as the distance between them, $R_{ij}=\left| \mathbf{R}_{i} - \mathbf{R}_{j} \right|$, increases. If a cut-off radius $r_{cut}$ is applied to the elements of $\mathbf{V}$, i.e., if we keep only the $V_{ij}$ satisfying $R_{ij}\leq r_{cut}$, then $\mathbf{V}$ becomes BB as well. The same simplification can be extended to all other quantities in Eqs.~(\ref{eq:system-solve}) to (\ref{eq:interaction}) so that only BB matrices similar to the one in Fig.~\ref{fig:device_to_bt_matrix} are involved. The accuracy of this truncation scheme has been validated by Deuschle et al.~\cite{deuschle2024_arXiv}.

\begin{figure*}[t]
    \centering
    \includegraphics[width=\textwidth]{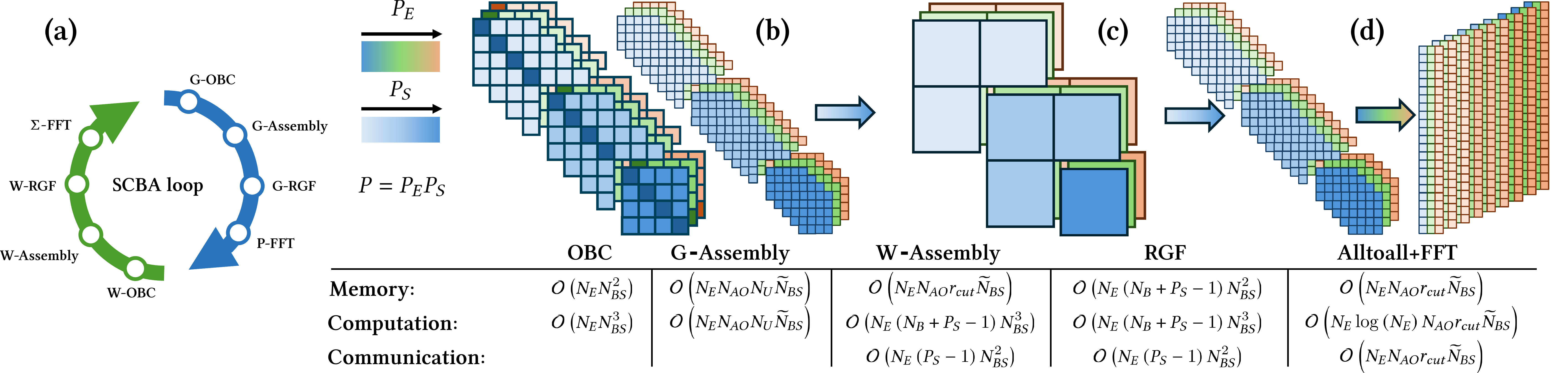}
    \vspace{-1em}
    \caption{Evolution of the data distribution throughout an SCBA iteration. (a) Illustration of the self-consistent $\mathbf{G}\rightarrow\mathbf{P}\rightarrow\mathbf{W}\rightarrow\mathbf{\Sigma}$ cycle. (b) Data distribution through the energy stack. (c) Selected solution of the quadratic matrix problem in Eq.~(\ref{eq:system-solve}) with RGF. (d) Data distribution through the non-zero elements of the matrices and FFT kernel.}
    \label{fig:quatrex_pipeline}
\end{figure*}

\subsection{Open Boundary Conditions}\label{sec:obc}

An overview of NEGF+sc$GW$ calculations and the SCBA loop is visualized in Fig.~\ref{fig:quatrex_pipeline}.
In the first step,
the OBC matrices are computed. It is assumed that the device of interest is contacted by two leads in thermodynamic equilibrium, placed at both extremities of the structure along the transport axis $x$. In case of transistors, these leads play the role of the source and drain electrodes. The OBC are then cast into the $\mathbf{B}_{OBC}(E)$ matrices in Eq.~(\ref{eq:system-solve}). They can be of retarded or lesser/greater type. Regardless, they occupy the upper left and lower right blocks of $\mathbf{B}_{OBC}(E)$ with size $N_{BS}\times N_{BS}$. All the other blocks are equal to zero. 

\subsubsection{Retarded OBC}
The physical properties of the leads give rise to a recursion relation with the following form \cite{Bowen1995}
\begin{equation}  \label{eq:retarded-recursion}
    \mathbf{x}^{R} = \left( \mathbf{m} - \mathbf{n} \mathbf{x}^{R}\mathbf{n}' \right)^{-1}.
\end{equation}
Here $\mathbf{x}^{R}$ denotes a surface matrix of size $N_{BS}\times N_{BS}$, whereas $\mathbf{m}$, $\mathbf{n}$ and $\mathbf{n}'$ refer to blocks of the $\mathbf{M}(E)-\mathbf{B}^{R}_{scatt}(E)$ matrices (see Fig. \ref{fig:device_to_bt_matrix}). 

Several methods have been proposed to solve the non-linear Eq.~(\ref{eq:retarded-recursion}). They can be generally divided into two classes, \textit{iterative} and \textit{direct} techniques. A straightforward iterative approach is to perform fixed-point iterations as
\begin{equation}
    \mathbf{x}^{R}_{i+1} = \left( \mathbf{m} - \mathbf{n} \mathbf{x}^{R}_{i}\mathbf{n}' \right)^{-1}
\label{eq:naive}    
\end{equation}
until $\left\|\mathbf{x}^{R}_{i+1} - \mathbf{x}^{R}_{i}\right\|$ falls below an imposed convergence criterion. Every iteration comes at the cost of multiplying three matrices and inverting one. Since the convergence (if any) of Eq.~(\ref{eq:naive}) can take 100s of iterations, alternatives have been developed to accelerate the process. The well-established Sancho-Rubio method \cite{sancho} significantly reduces the number of required iterations, but it still remains in the order of 10s. Also, as the convergence rate of Eq.~(\ref{eq:naive}) can be different for each energy and we want to treat the $N_E$ points in parallel, iterative methods potentially lead to load imbalance.

Another approach is to determine the $\mathbf{x}^{R}$ matrix in Eq.~(\ref{eq:retarded-recursion}) \textit{directly} by solving a polynomial eigenvalue problem (PEVP) \cite{luisierAtomisticSimulationNanowires2006, bruckEfficientAlgorithmsLargescale2017}
\begin{equation}
    \left[\sum_{i=-N_U}^{N_U} \lambda^{i} \widetilde{\mathbf{m}}_{i}\right]\phi = 0,
\label{eq:pevp}
\end{equation}
where the $\widetilde{\mathbf{m}}_i$ matrices are of size $\widetilde{N}_{BS}\times\widetilde{N}_{BS}$. They are also extracted from $\mathbf{M}(E)-\mathbf{B}^{R}_{scatt}(E)$. The surface function $\mathbf{x}^{R}$ can be reconstructed from the eigenvectors $\phi$ and eigenvalues $\lambda$.

Contour integral methods such as the Beyn algorithm can be harnessed to tackle Eq.~(\ref{eq:pevp}) \cite{beynIntegralMethodSolving2012}. A contour is first defined in the complex plane, two projectors $\mathbf{Q}_0$ and $\mathbf{Q}_1$ are created by solving sparse linear systems of equations at predefined points on this contour, the results are summed up, a singular value decomposition (SVD) of $\mathbf{Q}_0$ is performed, and the produced data is used to assemble a regular, non-symmetric eigenvalue problem (EVP) of significantly reduced dimensions. The EVP is solved to obtain the desired $\phi$, $\lambda$, and finally $\mathbf{x}^{R}$ \cite{bruckEfficientAlgorithmsLargescale2017}.  Once $\mathbf{x}^{R}$ is available, the non-zero blocks of $\mathbf{B}^{R}_{OBC}$ can be computed as $\mathbf{n} \mathbf{x}^{R}\mathbf{n}'$. While the Beyn algorithm returns accurate surface functions, it involves SVD and non-symmetric EVP operations that do not perform well on GPUs \cite{deuschle2024_sc}.

\subsubsection{Lesser/Greater OBC}
The lesser/greater boundary self-energy, $\mathbf{\Sigma}^{\lessgtr}_{OBC}$, can be derived analytically from the retarded surface Green's function $\mathbf{g}^{R}=\mathbf{x}^{R}$ in Eq.~(\ref{eq:retarded-recursion}) through the fluctuation-dissipation theorem \cite{fluctuation}. In case of the screened Coulomb interaction, the calculation of the required surface function $\mathbf{w}^{\lessgtr}$ is more complex as it involves solving a discrete-time Lyapunov equation \cite{deuschle2024_arXiv}
\begin{equation}\label{eq:lyapunov}
    \mathbf{w}^{\lessgtr} = \mathbf{q}^{\lessgtr} - \mathbf{a}\mathbf{w}^{\lessgtr}\mathbf{a}^{\dagger},
\end{equation} 
where $\mathbf{q}^{\lessgtr}$ and $\mathbf{a}$ are blocks of size $N_{BS}\times N_{BS}$ that can be extracted from the $\mathbf{P}$, $\mathbf{V}$, and $\mathbf{w^R}$ matrices. Equation (\ref{eq:lyapunov}) is standard in control systems, but not yet in quantum transport. As in the retarded case, it can be solved iteratively \cite{Poloni2020} or directly \cite{Kitagawa1977}. None of these techniques is ideal: The former converges slowly, and a matrix of size slightly smaller than $N_{BS}$ must be diagonalized in the latter.

\subsection{Selected System Solver}

The second step of the NEGF+sc$GW$ method is concerned with the ``selected'' solution of the quadratic matrix problem in Eq.~(\ref{eq:system-solve}) where both the $\mathbf{M}(E)$ (left-hand-side, (LHS)) and $\mathbf{B}^{\lessgtr}(E)$ (right-hand-side, (RHS)) matrices are block-banded as in Fig.~\ref{fig:device_to_bt_matrix}. ``Selected,'' in this context, means that we are not interested in all entries of the solution $\mathbf{X}^{\lessgtr}(E)$, but only in those that lie within the $r_{cut}$ boundaries in Fig.~\ref{fig:device_to_bt_matrix}. The so-called recursive Green's function (RGF) solver provides the desired elements of $\mathbf{X}^{\lessgtr}(E)$ with little fill-in \cite{rgf}. However, it does not apply to general BB matrices, but only to block-tridiagonal (BT) ones. To address this issue, $N_U$ primitive blocks $\mathbf{h}_{ij}$ in Fig.~\ref{fig:device_to_bt_matrix} are grouped together to form transport cells with green-colored borders and size $N_{BS} = \widetilde{N}_{BS} N_U$. This partitioning leads to a BT sparsity pattern with $N_B$ diagonal blocks. 

\subsubsection{Assembly of the Linear System}\label{sec:assembly}

Before solving Eq.~(\ref{eq:system-solve}) for a given energy point $E$, its $\widetilde{\mathbf{M}}$ and $\mathbf{B}^{\lessgtr}$ matrices must first be assembled (see Table \ref{tab:quantities}). More concretely, in the $\mathbf{G}$-solver, the matrix
\begin{equation}
    \widetilde{\mathbf{M}}(E) = E\mathbf{S}_{\mathbf{DFT}} - \mathbf{H}_{\mathrm{DFT}} - \mathbf{\Sigma}^R_{scatt}(E) - \mathbf{\Sigma}^R_{OBC}(E)
\end{equation}
has to be generated, which requires performing an element-wise subtraction of BB matrices and two OBC blocks. To assemble the RHS ($\Sigma^{\lessgtr}$), $\Sigma^{\lessgtr}_{scatt}$ and $\Sigma^{\lessgtr}_{OBC}$ must be added element-wise. The resulting computational cost is in the same order as the largest number of non-zeros among these matrices, i.e., $\mathcal{O}(N_{AO}\mathrm{max}(H_{BW},\Sigma_{BW}))$, where $H_{BW}$ ($\Sigma_{BW}$) is the bandwidth of $\mathbf{H}_{\mathbf{DFT}}$ ($\mathbf{\Sigma}$).

In the $\mathbf{W}$-solver, the construction of both $\widetilde{\mathbf{M}}$ and $\mathbf{B}^{\lessgtr}$ involves matrix multiplications, $\mathbf{V}\mathbf{P}^R$ and $\mathbf{V}\mathbf{P}^{\lessgtr}\mathbf{V}^{\dagger}$, respectively. Since $\mathbf{P}^{R,\lessgtr}$ and $\mathbf{V}$ have the same $r_{cut}$-dependent bandwidth $P_{BW}=V_{BW}$, the computational cost of assembling $\widetilde{\mathbf{M}}$ and $\mathbf{B}^{\lessgtr}$ is equal to $\mathcal{O}(N_{AO}P_{BW}^2)$. These operations can be performed either in a BB or even BT representation to increase the computational efficiency. In the latter case, the block-bandwidth of the matrices is three ($\mathbf{V}$, $\mathbf{P}$), five ($\mathbf{V}\mathbf{P}^R$), and seven ($\mathbf{V}\mathbf{P}^{\lessgtr}\mathbf{V}^{\dagger}$). The overall complexity increases to $\mathcal{O}(N_BN_{BS}^3)$, but the multiplication of larger blocks is ideally suited to GPUs.

\subsubsection{Recursive Green's Function (RGF) Solver}\label{sec:rgf}

\begin{figure}[h]
    \centering
    \includegraphics[width=\columnwidth]{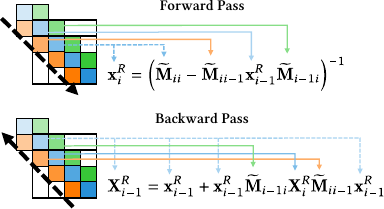}
    \vspace{-2em}
    \caption{Visualization of RGF's recursive Schur complement approach to obtain the main diagonal blocks of the selected inverse of $\widetilde{\mathbf{M}}$, $\mathbf{X}^R$. The forward pass iterates from the top left to the bottom right, while the backward pass iterates in the opposite direction. In every $i$-th forward pass iteration, we consider the $\widetilde{\mathbf{M}}_{i:,i:}$ submatrix and compute $\mathbf{x}^R_i$, as in Eq.~(\ref{eq:forward-pass}). Similarly, in the backward pass, the computation of $\mathbf{X}^R_{i-1}$ depends on the value from the previous iteration, $\mathbf{X}^R_{i}$ (Eq.~(\ref{eq:backward-pass})).}
    \label{fig:recursive-schur}
\end{figure}

In the device modeling community, the RGF algorithm is widely seen as the most powerful method to solve Eq.~(\ref{eq:system-solve}) for Green's functions \cite{rgf} and for screened Coulomb interactions \cite{deuschle2024_arXiv}. It relies on a recursive calculation of the Schur complement of the system matrix $\widetilde{\mathbf{M}}$ and exploits the BT sparsity pattern of this matrix and of $\mathbf{B}^{\lessgtr}$. The RGF method articulates itself around a forward pass during which intermediate quantities, $\mathbf{x}^R$ and $\mathbf{x}^{\lessgtr}$, are computed, followed by a backward pass that produces the targeted blocks of $\mathbf{X}^{\lessgtr}$ (diagonal and first off-diagonal).
One iteration of the forward and backward passes is shown in Fig.~\ref{fig:recursive-schur} to compute the selected inverse of $\widetilde{\mathbf{M}}$, $\mathbf{X}^R$.

The recursion starts, for example, at index $i=1$ by determining $\mathbf{x}^{R}_1=\widetilde{\mathbf{M}}_{11}^{-1}$ and $\mathbf{x}^{\lessgtr}_1=\mathbf{x}^{R}_1\mathbf{B}^{\lessgtr}_{11}\mathbf{x}^{R\dagger}_1$,
where all matrices are of size $N_{BS}\times N_{BS}$, i.e., they correspond to the green blocks in Fig.~\ref{fig:device_to_bt_matrix}. The RGF algorithm provides then a recursion for the indices $2\leq i\leq N_B$
\begin{eqnarray}
\mathbf{x}^{R}_i&=&\left(\widetilde{\mathbf{M}}_{ii}-\widetilde{\mathbf{M}}_{ii-1}\mathbf{x}^{R}_{i-1}\widetilde{\mathbf{M}}_{i-1i}\right)^{-1},\label{eq:forward-pass}\\
\mathbf{x}^{\lessgtr}_i&=&\mathbf{x}^{R}_i\left(\mathbf{B}^{\lessgtr}_{ii}+\widetilde{\mathbf{M}}_{ii-1}\mathbf{x}^{\lessgtr}_{i-1}\widetilde{\mathbf{M}}^{\dagger}_{i-1i}-\left[\mathbf{y}_{i}^{\lessgtr}-\mathbf{y}_{i}^{\lessgtr\dagger}\right]\right)\mathbf{x}^{R\dagger}_i,
\end{eqnarray}
with $\mathbf{y}^{\lessgtr}_i=\widetilde{\mathbf{M}}_{ii-1}\mathbf{x}^{\lessgtr}_{i-1}\mathbf{B}^{\lessgtr}_{i-1i}$. At the end of the forward pass, $\mathbf{X}^{\lessgtr}_{N_B}=\mathbf{x}^{\lessgtr}_{N_B}$ and $\mathbf{X}^{R}_{N_B}=\mathbf{x}^{R}_{N_B}$. This marks the beginning of the backward pass during which the index $i$ goes from $N_B-1$ to 1
\begin{eqnarray}
\mathbf{X}^{R}_i&=&\mathbf{x}^{R}_i+\mathbf{x}^{R}_i\widetilde{\mathbf{M}}_{ii+1}\mathbf{X}^{R}_{i+1}\widetilde{\mathbf{M}}_{i+1i}\mathbf{x}^{R}_i,\label{eq:backward-pass}\\
\mathbf{X}^{\lessgtr}_i&=&\mathbf{x}^{\lessgtr}_i+\mathbf{x}^{R}_i\widetilde{\mathbf{M}}_{ii+1}\mathbf{X}^{\lessgtr}_{i+1}\widetilde{\mathbf{M}}^{\dagger}_{i+1i}\mathbf{x}^{R\dagger}_i-\nonumber\\ &&\left(\mathbf{Y}_{i}^{\lessgtr}-\mathbf{Y}_{i}^{\lessgtr\dagger}\right)+
\left(\mathbf{Z}_{i}^{\lessgtr}-\mathbf{Z}_{i}^{\lessgtr\dagger}\right).
\end{eqnarray}
Here, $\mathbf{Y}_{i}^{\lessgtr}$ and $\mathbf{Z}_{i}^{\lessgtr}$ are defined as $\mathbf{Y}_{i}^{\lessgtr}=\mathbf{x}^{R}_i\mathbf{B}^{\lessgtr}_{ii+1}(\mathbf{x}^R_{i}\widetilde{\mathbf{M}}_{ii+1}\mathbf{X}^R_{i+1})^{\dagger}$ and $\mathbf{Z}_{i}^{\lessgtr}=\mathbf{x}^{R}_i\widetilde{\mathbf{M}}_{ii+1}\mathbf{X}^R_{i+1}\widetilde{\mathbf{M}}_{i+1i}\mathbf{x}^{\lessgtr}_{i}$, respectively. The first off-diagonal blocks of $\mathbf{X}^{\lessgtr}$ ($\mathbf{X}^{\lessgtr}_{ii\pm1}$) can also be computed with RGF \cite{deuschle2024_arXiv}.

There are two main challenges associated with the RGF algorithm. First, owing to its recursive nature, the method is inherently sequential and thus limited to moderately large device structures. Secondly, because of the accumulation of small numerical errors over multiple SCBA iterations, the $\mathbf{X}^{\lessgtr}$ matrices tend to lose their symmetry ($X^{\lessgtr}_{ij}=-X^{\lessgtr*}_{ji})$, which deteriorates the convergence of NEGF+sc$GW$. Regular symmetrizations of $\mathbf{X}^{\lessgtr}$ are therefore needed.  

\subsubsection{Computational Complexity}
The RGF algorithm consists of $N_B$ steps in the forward and backward pass. For each of this step, matrices of size $N_{BS}\times N_{BS}$ are either multiplied or inverted such that its computational complexity is equal to $\mathcal{O}(N_{B}N_{BS}^{3})$, or equivalently $\mathcal{O}(N_{AO}N_{BS}^{2})$. However, $N_E$ energy points must be computed for each SCBA iteration so that the cost increases to $\mathcal{O}(N_EN_{B}N_{BS}^{3})$ (see Table \ref{tab:competitors}). Without RGF, it would be $\mathcal{O}(N_EN_{AO}^{3})$.

\subsection{Energy Convolutions}
The Green's functions and screened Coulomb interactions enter Eq.~(\ref{eq:interaction}) to calculate either the scattering self-energy $\mathbf{\Sigma}(E)$ (element-wise multiplication of $\mathbf{G}(E')$ and $\mathbf{W}(E-E')$) or the polarization function $\mathbf{P}(E)$ (element-wise multiplication of two $\mathbf{G}$ at $E'$ and $E-E'$). The required energy convolutions can be replaced by fast Fourier transforms (FFTs), which provide efficient conversions between the energy and time domains (and vice versa), thereby reducing the complexity of Eq.~(\ref{eq:interaction}) from $\mathcal{O}(N_E^2)$ to $\mathcal{O}(N_E\log N_E)$.  

The computational complexity is not the only challenge here. It can be seen that the solution of Eq.~(\ref{eq:system-solve}) returns the desired $(i,j)$ entries of the $\mathbf{X}^{\lessgtr}$ matrix for one given energy $E$. All these energies are independent of each other and can be treated in parallel. However, in Eq.~(\ref{eq:interaction}), obtaining $\mathbf{B}_{scatt}(E)$ requires gathering all energies of two matrices $\mathbf{X}_1$ and $\mathbf{X}_2$ that are multiplied element-wise. Hence, the FFTs can be performed for specific $(i,j)$ entries, independently from the others. To address this data distribution duality, a solution was proposed in \cite{ziogas}: data transposition (see Fig. \ref{fig:quatrex_pipeline}). When dealing with Eq.~(\ref{eq:system-solve}), each rank stores all $(i,j)$ entries of $\mathbf{X}^{\lessgtr}$ for one (or a few) energy point(s). After data transposition, they store all energies of $\mathbf{X}_1$ and $\mathbf{X}_2$ for a few $(i,j)$ entries, and can then process Eq.~(\ref{eq:interaction}).

\subsection{Observables}
The simulation of nano-transistors such as the NRFET from Fig.~\ref{fig:ribbonfet} is expected to produce relevant observables, e.g., the local density-of-states (DOS), charge density $\rho(r)$, or electronic (energy) current $I_d$ ($I_{dE}$). All can be derived from the diagonal and first off-diagonal blocks of the lesser and greater Green's functions. The $GW$-renormalized band structure can also be obtained from the DFT Hamiltonian $\mathbf{H}_{\mathbf{DFT}}$ and the retarded scattering self-energy $\mathbf{\Sigma}^R_{scatt}$.

\section{Innovations Realized}

To address all the challenges arising in the NEGF+sc\textit{GW} method, we present algorithmic and programming developments that leverage both the physics at play and the underlying mathematical structure of the computational problem. 
The innovations and optimizations concern the programming model, symmetry exploitation, dynamic OBC memoization, and a distributed solver to enable unprecedented spatial domain decomposition of Eq.~(\ref{eq:system-solve}).

\subsection{Programming Model}

We implemented a new, modular NEGF+sc\textit{GW} simulator in Python, making use of the NumPy~\cite{harris2020array}, CuPy~\cite{cupy_learningsys2017}, mpi4py~\cite{dalcinMpi4pyStatusUpdate2021}, Numba~\cite{lam2015numba}, and Scipy\cite{2020SciPy-NMeth} packages, which provide a high-level interface to low-level, high-performing vendor-specific BLAS, LAPACK, FFT, and communication libraries. By harnessing the abstractions of these packages, our software is completely hardware-agnostic and runs on any CPU- and GPU-based system with different communication libraries (MPI / *CCL). Where package-provided implementations do not suffice, we implement and dispatch custom CPU and GPU kernels, using the just-in-time compilation functionality of CuPy and Numba. We also develop wrappers for certain linear algebra kernels (SVD, QR, EVP, etc.), to facilitate dispatching to the CPU.
This is needed in case GPU implementations do not exist (non-symmetric EVP) or are outperformed by CPU ones.

\subsection{Exploiting Symmetry}
\label{sec:symmetrization}

As mentioned in Section \ref{sec:rgf}, the lesser/greater quantities of the NEGF+sc$GW$ model must fulfill a very specific symmetry relationship ($X^{\lessgtr}_{ij}=-X^{\lessgtr*}_{ji}$). At each execution of the RGF algorithm, slight deviations from this condition are induced, for example, due to the floating-point arithmetic. These errors propagate to the $\mathbf{B}^{\lessgtr}_{scatt}$ matrices, which should also satisfy $B^{\lessgtr}_{scatt,ij}=-B^{\lessgtr*}_{scatt,ji}$. At the next SCBA iteration, they are injected back into the RGF solver. Overall, the errors accumulate over time, slowing down or even preventing the convergence of the $\mathbf{G}\rightarrow\mathbf{P}\rightarrow\mathbf{W}\rightarrow\mathbf{\Sigma}$ cycle.

By enforcing the desired symmetry properties at each SCBA iteration, convergence is accelerated, or restored in the worst-case scenario. The symmetrization needed is applied by computing $X^{\lessgtr}_{ij}=(X^{\lessgtr}_{ij}-X^{\lessgtr*}_{ji})/2$, which necessitates that both the upper and lower off-diagonal blocks of the matrix $\mathbf{X}^{\lessgtr}$ are determined and stored explicitly. The same must be done for $\mathbf{B}^{\lessgtr}$. Such an approach incurs significant computational and memory overhead. 

In this work, we improve the situation by absorbing the symmetrization into the data structure and the main computational kernels. Hence, symmetries are enforced on-the-fly, without sacrificing the convergence stability offered by computing all values.
The computational workload remains mostly unchanged, but the memory cost is significantly lowered by storing only the upper triangular part of the quantities obeying symmetry properties. Furthermore, the communication volume during data transposition and the time to calculate the matrices $\mathbf{B}^{\lessgtr}_{scatt}$ with Eq.~(\ref{eq:interaction}) are halved. 

\subsection{OBC Memoization}\label{sec:memoization}

In Section \ref{sec:obc}, we highlighted that direct solutions of the retarded and lesser/greater OBCs with Eqs.~(\ref{eq:retarded-recursion}) and (\ref{eq:lyapunov}), respectively, always provide accurate results, contrary to iterative solutions of the same equations, whose convergence is slow, especially if the initial guess for the quantity searched is set to zero. By closely analyzing the convergence behavior of the NEGF+sc\textit{GW} scheme for different device examples and through many SCBA cycles, we found that, already after relatively few iterations, the retarded and lesser/greater OBC blocks tend to stabilize, i.e., they stop varying significantly from one SCBA iteration to the next. This observation motivated us to switch from a direct to an iterative approach as soon as the OBC blocks become stable. Indeed, the time to iteratively compute these blocks drastically decreases when the solution from the previous SCBA iteration is very close to the new one and is used as an initial guess. Moreover, our code identifies the moment to dynamically transition from one mode of calculation to the other.

The switching from a direct to an iterative method requires storing the OBC blocks at each SCBA iteration. This technique, therefore, trades computational speed-up for increased memory usage, which lends itself well after having previously ``freed up'' memory by exploiting data symmetry.
We thus implemented a memoization scheme for the retarded and lesser/greater OBC blocks. More concretely, we initially cache the $\mathbf{x}^{R\lessgtr}(E)$ obtained from a call to the corresponding \textit{direct} OBC solver for each contact (left and right) in each subsystem ($\mathbf{G}$ / $\mathbf{W}$). In the following SCBA iteration, we retrieve the cached results and compute the update $\|\mathbf{x}^{R\lessgtr}_{i+1} - \mathbf{x}^{R\lessgtr}_{i}\|$. From this update, the memoizer estimates whether it can reach convergence in a predetermined number of fixed-point iterations $N_{FPI}$. If a predefined condition is satisfied, the $N_{FPI}$ iterations are performed, the result is returned, and the cache is updated. If the allotted $N_{FPI}$ is deemed to be insufficient to achieve convergence, the direct solver is called instead. In this way, the memoizer decides at runtime whether to use the more robust direct method or the significantly more performant iterative one.

Since we solve the OBC on all ranks, the speed-up only truly manifests if all ranks decide to memoize. Allotting a fixed number of iterations $N_{FPI}$ prevents load imbalance, should the memoizers on all ranks succeed. In our analysis, we found that the lesser/greater recursion relation in Eq.~(\ref{eq:lyapunov}) typically stabilizes within fewer than 10 iterations. The retarded recursion relation in Eq.~(\ref{eq:retarded-recursion}) also tends to stop varying much after 20 iterations. Since typically 100s of SCBA iterations are needed to reach convergence, this means we usually can expect all memoizers to succeed in >90\% of iterations.

\subsection{Spatial Domain Decomposition}\label{sec:domain_decomposition}

The computation of OBCs and the subsequent assembly of each subsystem's matrices are embarrassingly parallel through the energy grid. Decomposition of the spatial domain, on the other hand, is significantly less straightforward due to the sequential dependence that the RGF algorithm introduces between spatial subdomains (see Section \ref{sec:rgf}). To overcome this limitation, we developed and integrated a novel distributed algorithm for the production of selected elements of the solution to Eq.~(\ref{eq:system-solve}) \cite{maillouSerinvScalableLibrary2025}. Similarly to RGF, this method is based on the Schur complement. Our algorithm extends prior works~\cite{pdiv,psr}, employing a nested-dissection scheme to permute the system matrix and the RHS $\mathbf{B}^{\lessgtr}$, enabling the concurrent solution of the equation in different sections of the spatial domain.

\begin{figure}[h]
    \centering
    \includegraphics[width=\columnwidth]{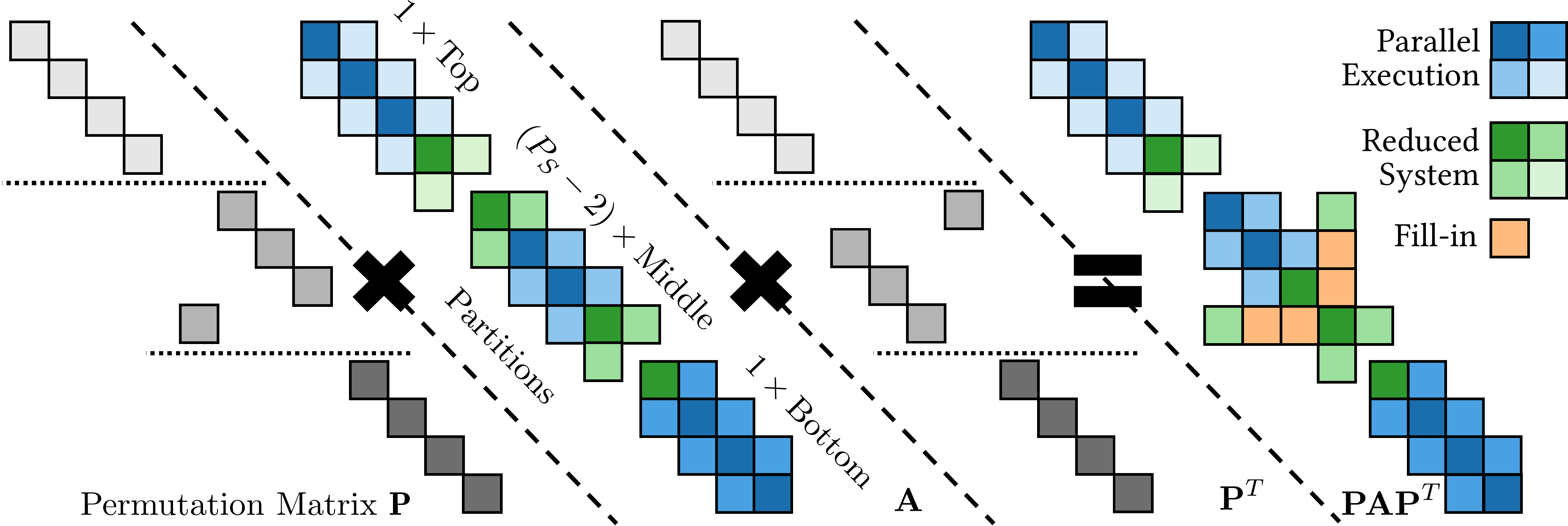}
    \vspace{-2em}
    \caption{Visualization of the nested-dissection scheme for RGF. From left to right: permutation matrix $\mathbf{P}$, BT matrix $\mathbf{A}$ (applies to both $\widetilde{\mathbf{M}}$ and $\mathbf{B}^\lessgtr$), $\mathbf{P}^T$, and $\mathbf{P}\mathbf{A}\mathbf{P}^T$. The scheme splits the matrices into $P_S$ partitions, 1 top, 1 bottom, and $P_S-2$ middle ones. The permutation applies separately to each partition, breaking the sequential dependencies among them in RGF's forward and backward passes, thus enabling parallel execution. Between the passes, a reduced system must be solved (green), inducing additional workload.}
    \label{fig:nested-dissection}
\end{figure}

The nested-dissection scheme is visualized in Fig.~\ref{fig:nested-dissection}.
A BT matrix $\mathbf{A}$ ($\widetilde{\mathbf{M}}$ and $\mathbf{B}^\lessgtr$) is split into $P_S$ disjoint partitions, as shown with different shades of blue.
We explicitly select the visualized \textit{arrow} partitioning so that every $\mathbf{A}_{i,j}$ off-diagonal block lies in the same partition as its symmetric $\mathbf{A}_{j,i}$.
The symmetrization techniques discussed in Section~\ref{sec:symmetrization} naturally extend to this spatial decomposition scheme without inducing extra communication across the partitions' boundaries.
Multiplying $\mathbf{A}$ with a suitable permutation matrix $\mathbf{P}$ per partition from the left and right exposes parallel sections, where RGF's forward and backward passes can be executed concurrently.
However, this parallelization comes at a cost: At the end of the forward pass, a \textit{reduced system} (in green color) must be constructed and selectively solved before proceeding with the backward pass.
Furthermore, each \textit{middle} partition has to compute additional $O(N_B/P_S)$ blocks as fill-in (in orange color), causing load imbalance.
The reduced system increases the total computational workload by $O(P_SN_{BS}^3)$ and the overall memory requirements by $O(P_SN_{BS}^2)$, while adding a $O(P_SN_{BS}^2)$ communication cost to gather the system, compared to sequential RGF. These additional costs can nevertheless be distributed over multiple ranks by applying the nested-dissection scheme to the reduced system recursively.

To take advantage of this nested-dissection solver, the domain decomposition must be applied to all other tasks as well. While this is trivial for the most part, the BT matrix multiplications in the assembly of $\widetilde{\mathbf{M}}$ and $\mathbf{B}^{\lessgtr}$ for $\mathbf{W}$ (see Section \ref{sec:assembly}) requires the implementation of a specific halo communication scheme to exchange blocks at the partition boundaries.

\begin{table*}[h]
\caption{
Workload, time, and performance per SCBA iteration on a single GCD (Frontier) / GPU (Alps) of the main kernels for the NW-1, NW-2, NR-16, and NR-23 device structures. Median values of at least 10 measurements are reported.}
\vspace{-1.1em}
\footnotesize
\centering
\begin{tabular}{l  c|c c c c ||c|c c c c || c c c c}
  \multicolumn{15}{c}{\normalsize \textbf{Devices \& Machines} \footnotesize} \\
    \vspace{-0.9em}
\\
\hline
\multicolumn{1}{r||}{Software} & \multicolumn{1}{c|}{\scriptsize{QuaTrEx$_{24}$}} & \multicolumn{4}{c||}{\textbf{This Work}} &  \multicolumn{1}{c|}{\scriptsize{QuaTrEx$_{24}$}} &\multicolumn{4}{c||}{\textbf{This Work}} &\multicolumn{4}{c}{\textbf{This Work}}\\
\hline
\multicolumn{1}{r||}{Device} &  \multicolumn{5}{c||}{NW-1} & \multicolumn{5}{c||}{NW-2} & \multicolumn{2}{c}{NR-16} & \multicolumn{2}{c}{NR-23}\\
\hline
\multicolumn{1}{r||}{Machine} & LUMI & \multicolumn{2}{c}{Frontier} & \multicolumn{2}{c||}{Alps} &LUMI & \multicolumn{2}{c}{Frontier} & \multicolumn{2}{c||}{Alps} & \multicolumn{2}{c}{Frontier} & \multicolumn{2}{c}{Alps}\\
\hline
\multicolumn{1}{r||}{Energies} & 32 & \multicolumn{2}{c}{50} & \multicolumn{2}{c||}{80} & 1 & \multicolumn{2}{c}{4} & \multicolumn{2}{c||}{6} &  \multicolumn{2}{c}{1} & \multicolumn{2}{c}{1}\\
 \multicolumn{1}{r||}{Memoizer} & \xmark & \xmark & \cmark & \xmark & \cmark  & \xmark &  \xmark & \cmark  & \xmark & \cmark  & \xmark & \cmark  & \xmark & \cmark\\
\hline
\vspace{-.6em}
\\
  \multicolumn{15}{c}{\normalsize \textbf{Workload [Tflop]} \footnotesize} \\
    \vspace{-0.9em}
\\
\hline
\multicolumn{1}{l||}{G: OBC} & 1.242 & 1.635 & 0.619 & 1.796 & 0.811 & -- & 8.554 & 4.878 & 12.831 & 7.317  & 9.686 & 5.809 & 9.686 & 5.809 \\
\multicolumn{1}{l||}{G: RGF} & 16.035 & \multicolumn{2}{c}{21.071} & \multicolumn{2}{c||}{28.821} & -- & \multicolumn{2}{c}{149.930} & \multicolumn{2}{c||}{224.895}  & \multicolumn{2}{c}{167.704} & \multicolumn{2}{c}{244.077}  \\
\hline
\multicolumn{1}{l||}{W: Assembly} &  &  & &  & &  &  & &  & &  &  & &  \\
\multicolumn{1}{l||}{\quad Beyn} & -- & 6.585 & 4.729 & 8.613 & 6.461 & -- & 6.556 & 4.878 & 9.834 & 7.317  & 7.629 & 5.809 & 7.629 & 5.809 \\
\multicolumn{1}{l||}{\quad Lyapunov} & -- & 8.613 & 5.900 & 11.114 & 10.964 & -- & 12.490 & 8.126 & 18.735 & 12.189  & 8.486 & 5.875 & 8.486 & 5.875 \\
\multicolumn{1}{l||}{\quad LHS-$\mathbf{B}_{\textit{scatt}}^R$} & -- & \multicolumn{2}{c}{5.170} & \multicolumn{2}{c||}{7.114} & -- & \multicolumn{2}{c}{36.292} & \multicolumn{2}{c||}{54.438 } & \multicolumn{2}{c}{44.287} & \multicolumn{2}{c}{64.504} \\
\multicolumn{1}{l||}{\quad RHS-$\mathbf{B}_{\textit{scatt}}^{\lessgtr}$} & -- & \multicolumn{2}{c}{21.076} & \multicolumn{2}{c||}{38.051} & -- & \multicolumn{2}{c}{149.518} & \multicolumn{2}{c||}{224.277} & \multicolumn{2}{c}{181.056} & \multicolumn{2}{c}{261.904} \\

\multicolumn{1}{l||}{\hquad Total} & 12.680 & 41.444 & 36.875 & 64.891 & 62.590 & -- & 204.856 & 198.814 & 307.284 & 298.221 & 241.458 & 237.027 & 342.523 & 338.092 \\
\hline
\multicolumn{1}{l||}{W: RGF} & 55.424 & \multicolumn{2}{c}{77.011}  & \multicolumn{2}{c||}{109.384} & -- & \multicolumn{2}{c}{149.90}  & \multicolumn{2}{c||}{224.895} & \multicolumn{2}{c}{167.704}  & \multicolumn{2}{c}{244.077} \\
\multicolumn{1}{l||}{Other} & -- & 4.643 & 6.706 & 10.611 & 13.898 & -- & 39.264 & 36.65 & 58.896 & 54.975 & 3.345 & 1.338 & 3.345 & 1.338 \\
\hline\hline
\multicolumn{1}{l||}{Total Work} & 85.381 & 145.804 & 142.283 & 215.500 & 209.603 & 151.069 & 552.534 & 540.202 & 828.801 & 810.303 & 589.897 & 579.582 & 843.707 & 833.392 \\
\hline
\vspace{-.6em}
\\
  \multicolumn{15}{c}{\normalsize \textbf{Time [s]} \footnotesize} \\
  \vspace{-0.9em}
\\
\hline
\multicolumn{1}{l||}{G: OBC} & 1.298 & 0.901 & 0.161 & 0.611 & 0.176 & 4.178 & 2.505 & 0.649 & 1.987 & 0.470 & 2.402 &0.650 & 2.371 & 0.250\\
\multicolumn{1}{l||}{G: RGF}   & 1.050 & \multicolumn{2}{c}{1.621} & \multicolumn{2}{c||}{1.396} & 2.445 & \multicolumn{2}{c}{8.402} & \multicolumn{2}{c||}{7.367} & \multicolumn{2}{c}{7.873} & \multicolumn{2}{c}{5.963} \\
\hline
\multicolumn{1}{l||}{W: Assembly} &  &  & &  & &  &  & &  & &  &  & &  \\
\multicolumn{1}{l||}{\quad Beyn} & 1.630 & 1.193 & 0.421& 1.311 & 0.745 & 1.885 & 2.376 & 0.581 & 1.890 & 0.423 & 1.992 & 0.533 & 2.237 & 0.208\\
\multicolumn{1}{l||}{\quad Lyapunov} & 1.098 & 4.468 & 0.372 & 2.211 & 0.264 & 2.628 & 11.713 & 0.457 & 13.008 &  0.378 & 19.658 & 0.501 & 22.578 & 0.279\\
\multicolumn{1}{l||}{\quad LHS-$\mathbf{B}_{\textit{scatt}}^R$}  & \multirow{2}{*}{3.547} & \multicolumn{2}{c}{0.384} & \multicolumn{2}{c||}{0.299} & \multirow{2}{*}{2.362}  & \multicolumn{2}{c}{1.539} & \multicolumn{2}{c||}{1.113}  & \multicolumn{2}{c}{1.672} & \multicolumn{2}{c}{1.257} \\
\multicolumn{1}{l||}{\quad RHS-$\mathbf{B}_{\textit{scatt}}^{\lessgtr}$} &   & \multicolumn{2}{c}{1.333} & \multicolumn{2}{c||}{0.916} & & \multicolumn{2}{c}{5.678} & \multicolumn{2}{c||}{4.3715} & \multicolumn{2}{c}{6.077} & \multicolumn{2}{c}{4.792}  \\

\multicolumn{1}{l||}{\hquad Total} & 6.230 & 7.378 & 2.51 & 4.7365 & 2.224 & 10.593 & 21.306 & 8.255 & 20.383 & 6.286 & 29.399  & 8.783 & 30.864 & 6.536 \\
\hline
\multicolumn{1}{l||}{W: RGF} & 2.866 & \multicolumn{2}{c}{3.705} & \multicolumn{2}{c||}{3.753} & 2.357 & \multicolumn{2}{c}{7.2105} & \multicolumn{2}{c||}{6.052} & \multicolumn{2}{c}{7.800} & \multicolumn{2}{c}{5.956} \\
\multicolumn{1}{l||}{Other} & 1.104 & \multicolumn{2}{c}{1.747} & \multicolumn{2}{c||}{1.1} & 6.244 & \multicolumn{2}{c}{~2.883} & \multicolumn{2}{c||}{1.088} & \multicolumn{2}{c}{4.556} & \multicolumn{2}{c}{1.021}  \\
\hline\hline
\multicolumn{1}{l||}{Total Time} & 12.458 & 15.376 & 9.744 &  11.769 & 8.643 & 25.817 & 42.097 & 27.404 & 36.943 &  21.263 & 52.695 &  29.662 & 46.175 &  19.726 \\
\hline\hline
\multicolumn{1}{l||}{Time/Energy} & 0.389 & 0.308 & \textbf{0.195} &  0.147 & \textbf{0.108} & 25.817 & 10.524 & \textbf{6.851} & 6.157 &  \textbf{3.544} & 52.695 &  \textbf{29.662} & 46.175 &  \textbf{19.726} \\
\hline
\vspace{-.6em}
\\
\multicolumn{15}{c}{\normalsize \textbf{Performance [Tflop/s]} \footnotesize} \\
  \vspace{-0.9em}
\\
\hline
\multicolumn{1}{l||}{Performance} & 6.854 & 9.483 & 14.602 & 18.311 & 24.251 & 5.852 & 13.125 & 19.713 & 22.435 & 38.109 & 11.195 & 19.540 & 18.272 & 42.248 \\
\hline
\label{tab:tflop_time_flops}
\end{tabular}
\end{table*}

\section{How Performance was Measured}

\subsection{Hardware and Measurement Setup}

Our measurements were done on the Alps and Frontier supercomputers. Alps is made of HPE Cray EX254n blades, each housing two nodes, for a total of 2,600~\cite{hoefler_alps}.
Every node contains four NVIDIA GH200 superchips (72-core Grace CPU and one Hopper GPU), with 96\;GB of high-bandwidth memory (HBM) each.
Intranode communication uses NVLink interconnect with 150\;GB/s bidirectional bandwidth per link.
Every GH200 is connected to a Slingshot network through a separate network interface card (NIC), with 25\;GB/s bidirectional bandwidth.
Hopper's theoretical maximum double-precision floating-point (tensor-core) performance (FP64) is 67\;Tflop/s.
Alps' Rpeak and Rmax~\cite{top500} per GH200 superchip are 55.3 and 41.8\;Tflop/s, respectively.
Frontier consists of 9,604 Cray EX 235a nodes~\cite{Atchley2023_frontier}.
Every node has four AMD Instinct MI250X GPUs and an AMD EPYC 64-core Trento CPU.
Each MI250X GPU is made of two separate graphics compute dies (GCD) with 64\;GB HBM memory.
Intranode communication uses Infinity Fabric interconnect with up to 50\;GB/s bidirectional bandwidth.
Similarly to Alps, each GPU is connected to a Slingshot network through a separate NIC with 25\;GB/s bidirectional bandwidth.
Each MI250X GPU (GCD) has a theoretical maximum FP64 (tensor-core) performance of 95.7 (47.9)\;Tflop/s.
Frontier's Rpeak and Rmax per GPU (GCD) are 53.5 (26.8) and 35.2 (17.6)\;Tflop/s, respectively.

\subsection{Test Structures}
We use, in total, eight different nanowire and nanoribbon devices, as presented in Table~\ref{tab:structure_param}, to benchmark the performance of our NEGF+sc$GW$ solver, QuaTrEx. All structures model a silicon transistor, the surface of which is passivated with hydrogen atoms. The first two (NW-1 and NW-2) are nanowires that are identical to the configurations labeled ``medium'' and ``large'' in \cite{deuschle2024_sc}. 
The six other devices, NR-\{16, 24, 40\} on Frontier and NR-\{23, 44, 80\} on Alps, are NRFETs with the same cross section ($\sim$1.5$\times$5 nm$^2$) as in Fig.~\ref{fig:ribbonfet}. They correspond to devices fabricated by Intel and reported in December 2024 at the International Electron Devices Meeting (IEDM) \cite{intelnr}.
Each transport cell of length $L_{TC}$=2.172 nm contains 1,056 atoms ($N_{BS}$=3,408).
The digits following ``NR-'' indicate the total number of transport cells.
Hence, structure NR-40, measures $L_{tot}$=86.9 nm comprising 42,240 atoms.

We use as many energy points per compute unit as possible, saturating the available GPU memory when running experiments at scale.
Due to GH200's larger HBM (96\;GB) compared to the MI250X GCD (64\;GB), the NW-X devices can be run with more energy points per compute device. Similarly, without domain decomposition, the largest possible structure on Alps is NR-23, while Frontier is limited to the smaller NR-16. When turning on domain decomposition onto two compute units ($P_S=2$), the maximum size increases to NR-24 on Frontier and NR-44 on Alps. Finally, with four compute units ($P_S =4$), we can run NR-40 on Frontier and NR-80 on Alps.

\subsection{Testing Methodology}
In our benchmarks, we report the median of at least 10 measurements obtained by executing consecutive SCBA iterations.
We set the self-energy to zero at the beginning of each iteration to ensure stability with a small number of nodes/energy points.
The first iteration is always discarded since it includes the just-in-time (JIT) compilation of compute kernels and other initial overhead. 

Workload measurements (all in FP64) are performed with the vendor-provided tools, AMD ROCProfiler (rocprof) and NVIDIA Nsight Compute (NCU).
NCU could only be used for NW-1 since its runtime drastically increases with the device size.
On NW-2, we interpolate the rocprof's measurements by scaling the relevant workload with the increase in the energy points.
On the NR-X configurations, we use rocprof's measurements for the kernels that are independent of the device length (e.g., OBCs).
On the other hand, the workload of the kernels that scale with device length (e.g., RGF) is dominated by BLAS level 3 calls (mainly GEMM).
Therefore, we individually profile those calls with rocprof and NCU, and subsequently multiply the result by their exact count.

\section{Performance Results}

\subsection{Micro-benchmarks}

To evaluate our software, QuaTrEx, and compare it with its predecessor QuaTrEx$_{24}$~\cite{deuschle2024_sc} (see Table \ref{tab:competitors}), we first discuss single-device (GPU Alps / GCD Frontier) performance of the main computational kernels, as shown in Table~\ref{tab:tflop_time_flops}. 
QuaTrEx$_{24}$'s measurements were obtained on LUMI~\cite{lumi}, which has the same architecture as Frontier, and are also reported per GCD.
Thanks to computational optimizations and memory savings attained by exploiting data symmetries, our software achieves a higher throughput of energy points per time.
We fit 1.56$\times$\;(NW-1) / 4.00$\times$\;(NW-2) the number of energies onto one MI250X GCD while simultaneously increasing the total runtime by only 1.23$\times$\;(NW-1) and 1.63$\times$\;(NW-2), for an overall 1.27$\times$\;(NW-1) and 2.45$\times$\;(NW-2) speed-up per energy.
The performance is further improved when enabling memoization, resulting in 2.00$\times$\;(NW-1) and 3.77$\times$\;(NW-2) speed-up per energy.
When simulating the NR-16 (Frontier) and NR-23 (Alps), our software achieves high performance, especially when OBC memoization is enabled; approximately 72.9\% and 76.4\% of the respective machine's Rpeak.

We proceed with the last four device configurations, NR-\{24, 40\} (Frontier) and NR-\{44, 80\} (Alps), containing an unprecedented number of atoms, 25,344, 42,240, 44,464, and 84,480, respectively.
To simulate such large devices, we employ spatial-domain decomposition with $P_S=2$ or $4$ processes treating a single energy.
Table~\ref{tab:intel-dist-tflop} presents our software's performance with the minimum number of compute devices required to run a simulation (approximately 2-4 energies).
As mentioned in Section~\ref{sec:domain_decomposition}, distributed RGF induces an increased workload for the middle partitions.
In this work, we do not employ any load-balancing techniques yet, so that the boundary partitions perform about 60\% of the middle partitions' workload. 

\begin{table}[h]
\footnotesize
\centering
\caption{Workload, time, and performance measurements for one energy point of the NR-\{24, 40, 44, 80\} devices, with spatial domain decomposition and $P_S=2$ or $4$.
Median values of at least 10 measurements are reported.}
\vspace{-1em}
\begin{tabular}{ l c c c c c }
\hline
\multicolumn{1}{r||}{Software} & \multicolumn{4}{c}{This Work} \\
\hline
\multicolumn{1}{r||}{Machine}  & \multicolumn{2}{c}{Frontier} & \multicolumn{2}{c}{Alps} \\
\hline
\multicolumn{1}{r||}{Device}  & NR-24 & NR-40 & NR-44 & NR-80 \\
\multicolumn{1}{r||}{$P_S$} & 2 & 4 & 2 & 4 \\
\multicolumn{1}{r||}{Energies}  & \multicolumn{4}{c}{1} \\
\multicolumn{1}{r||}{Memoizer}  & \cmark & \cmark &\cmark & \cmark \\
\hline
\vspace{-.6em}
\\
\multicolumn{5}{c}{\small \textbf{Workload [Tflop]}\footnotesize} \\
    \vspace{-0.9em}
\\
\hline
\multicolumn{1}{l||}{Top partition}  & 483.547 & 490.711 & 899.501 & 906.632 \\
\multicolumn{1}{l||}{Middle partitions (per rank)}   & --- & 771.766 & --- & 1,536.419 \\
\multicolumn{1}{l||}{Bottom partition}  & 526.531 & 532.392 & 948.770 & 954.605 \\
\multicolumn{1}{l||}{Total} & 1,010.078 & 2,566.635 & 1,848.271 & 4,934.075 \\
\hline
\vspace{-.6em}
\\
\multicolumn{5}{c}{\small \textbf{Total Time [s]}\footnotesize} \\
    \vspace{-0.9em}
\\
\hline
\multicolumn{1}{l||}{All partitions} & 25.452 & 34.543 & 21.742 & 36.722 \\
\hline
\vspace{-.6em}
\\
\multicolumn{5}{c}{\small \textbf{Performance [Tflop/s]}\footnotesize} \\
    \vspace{-0.9em}
\\
\hline
\multicolumn{1}{l||}{Top partition}  & 18.998 & 14.206 & 41.372 & 24.689 \\
\multicolumn{1}{l||}{Middle partitions (per rank)}   & --- & 22.342 & --- & 41.839\\
\multicolumn{1}{l||}{Bottom partition}  & 20.687 & 15.412 & 43.638 & 25.995 \\
\multicolumn{1}{l||}{Total}  & 39.686 & 74.303 & 85.009 & 134.363 \\
\hline
\label{tab:intel-dist-tflop}
\end{tabular}
\end{table}

\begin{figure*}[h]
    \centering
    \includegraphics[width=\textwidth]{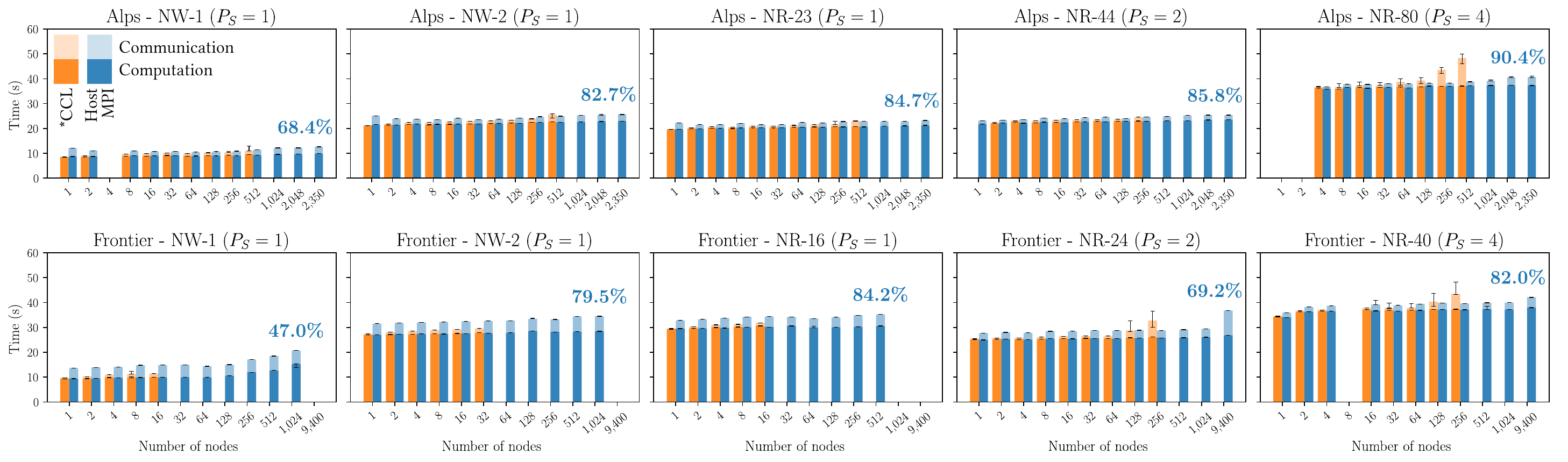}
    \vspace{-2em}
    \caption{Weak scaling as a function of the number of energy points $N_E$. Reporting median runtimes for two different communication backends: *CCL (orange) and Host MPI (blue). The runtimes are further divided into two parts: communication (lighter shade) and computation (darker shade). We compute median values of $\sim$10 measurements and 95\% CI (error bars) for a single SCBA iteration of the NW-1 and NW-2 nanowires (Alps and Frontier), as well as for the NR-\{23,~44,~80\} (Alps) and NR-\{16,~24,~40\} (Frontier) configurations.}
    \label{fig:weak_scaling}
\end{figure*}

\subsection{Weak Scaling}

We perform weak scaling experiments on the number of energy points $N_E$.
Except for the FFT, the workload scales linearly with $N_E$.
We use one (MPI) rank per GPU (Alps) or GCD (Frontier).
The number of energy points per rank for NW-1, NW-2, NR-16, and NR-23 is shown in Table~\ref{tab:tflop_time_flops}.
Spatial-domain decomposition is used only for NR-\{24, 44\}, with $P_S=2$, and NR-\{40, 80\}, with $P_S=4$ ranks treating a single energy point.
The computation, communication, and total runtimes per SCBA iteration with the memoizer enabled are presented in Fig.~\ref{fig:weak_scaling}.
The data points are median runtimes, and the error bars are 95\% CI computed with bootstrapping~\cite{bootstrapping}.
We annotate the achieved parallel efficiency at the maximum number of nodes for every machine/device combination.

Our software supports different communication libraries: *CCL, GPU-aware MPI, and ``host MPI'' (copying data to the host).
We show only *CCL and host MPI, as they performed the best.
On Alps, the best network performance up to 256-512 nodes is achieved with NCCL, after which it exhibits instability and slowdowns, necessitating a switch to host MPI.
On Frontier, we observe similar performance with RCCL, but it starts to become unstable earlier, at about 32 nodes.
Communication induced by spatial-domain decomposition is intranode, and therefore handled with *CCL.

In all experiments, we achieve a flat scaling until 128 nodes. 
Afterward, due to FFT's computational cost and other delays resulting in compute time increases, the smaller NW-1 device loses scaling efficiency.
The rest of the devices continue to scale well up to 1024 nodes or even the full scale.
When running almost at the full scale of Alps, we observe significant spurious delays in compute kernels, more precisely in the system solvers, but only for a subset of the executed iterations.
We attribute those to faulty GPUs, but we were unable to repeat the experiments due to limited resources.

In Table~\ref{tab:intel-dist-hero}, we summarize our software's performance at almost the full scale of Frontier and Alps for the NR-\{24, 40\} and NR-\{44-80\} NRFETs, respectively.
On the larger machine, Frontier, we surpass 1 Eflop/s FP64 sustained performance on 9,400 nodes, achieving with NR-40 57.0\% and 86.5\% of Frontier's Rpeak and Rmax (scaled for number of nodes).

\begin{table}[h]
\footnotesize
\centering
\caption{Large-scale simulations on Alps and Frontier.}
\vspace{-1em}
\begin{tabular}{ l c c c c c }
\hline
\multicolumn{1}{l||}{Machine}  & \multicolumn{2}{c|}{Frontier} & \multicolumn{2}{c}{Alps} \\
\multicolumn{1}{l||}{Nodes}  &  \multicolumn{2}{c|}{9,604}  &  \multicolumn{2}{c}{2,600} \\
\multicolumn{1}{l||}{Rmax [Pflop/s] }  &  \multicolumn{2}{c|}{1,353.00} &  \multicolumn{2}{c}{434.90} \\
\multicolumn{1}{l||}{Rpeak [Pflop/s] }  &  \multicolumn{2}{c|}{2,055.72} & \multicolumn{2}{c}{574.84} \\
\hline
\vspace{-.5em}
\\
\hline
\multicolumn{1}{l||}{Device}  &  NR-24 & \multicolumn{1}{c|}{NR-40} & NR-23 & NR-44 \\
\multicolumn{1}{l||}{$P_S$}  &  2 & \multicolumn{1}{c|}{4} & 1 & 2 \\
\multicolumn{1}{l||}{Memoizer} & \cmark & \multicolumn{1}{c|}{\cmark} & \cmark & \cmark \\
\hline
\hline
\multicolumn{1}{l||}{Atoms}  & 25,344  & \multicolumn{1}{c|}{\textbf{42,240}} & 24,288 & 46,464 \\
\multicolumn{1}{l||}{Total Energies}  & 37,600 & \multicolumn{1}{c|}{18,800} & 9,400 & 4,700 \\
\multicolumn{1}{l||}{Nodes (\#N)}  & \multicolumn{2}{c|}{\rule[1.5pt]{0.7cm}{0.4pt}~9,400~\rule[1.5pt]{0.7cm}{0.4pt}} & \multicolumn{2}{c}{\rule[1.5pt]{0.7cm}{0.4pt}~2,350~\rule[1.5pt]{0.7cm}{0.4pt}} \\
\multicolumn{1}{l||}{GCDs / GPUs}  & \multicolumn{2}{c|}{\rule[1.5pt]{0.7cm}{0.4pt}~75,200~\rule[1.5pt]{0.7cm}{0.4pt}} & \multicolumn{2}{c}{\rule[1.5pt]{0.7cm}{0.4pt}~9,400~\rule[1.5pt]{0.7cm}{0.4pt}}  \\
\multicolumn{1}{l||}{Workload [Pflop]}  & 37,978.933 & \multicolumn{1}{c|}{48,252.738} & 7,833.885 & 8,686.874 \\
\multicolumn{1}{l||}{Time per iteration [s]}  & 36.789 & \multicolumn{1}{c|}{\textbf{42.104}} & 23.286 & 25.353 \\
\multicolumn{1}{l||}{Performance [Pflop/s]}  &  1,032.345 & \multicolumn{1}{c|}{\textbf{1,146.037}} & 336.420 & 342.637 \\
\multicolumn{1}{l||}{Scaling efficiency [\%]}  &  69.2 & \multicolumn{1}{c|}{82.0} & 84.7 & 85.8 \\
\multicolumn{1}{l||}{Rmax (\#N scaled) [\%] }  & 76.3 (80.0) & \multicolumn{1}{c|}{\textbf{84.7} (86.5)} & 77.4 (85.6) & 78.8 (87.2) \\
\multicolumn{1}{l||}{Rpeak (\#N scaled) [\%] } &  50.2 (51.3) & \multicolumn{1}{c|}{55.7 (57.0)} & 58.5 (64.8) & 59.6 (65.9) \\
\hline
\label{tab:intel-dist-hero}
\vspace{-1em}
\end{tabular}
\end{table}

\section{Implications}
\label{sec:discussion}

We presented a new \textit{ab-initio} quantum transport solver called \mbox{QuaTrEx} based on the atomistic non-equilibrium Green's function formalism and capable of modeling electron-electron interactions within the $GW$ approximation in transistor structures with dimensions comparable to experimental devices \cite{intelnr}. 
Compared to the state of the art from 2024, we have increased the maximum achievable simulation workload, proportional to $O(N_EN_BN_{BS}^3)$, by a factor of almost 16 (18,800 vs. 14,400 for $N_E$, 40 vs. 16 for $N_B$, 3,408 vs. 2,016 for $N_{BS}$), while increasing the walltime by only 35\% (42.1 s vs. 31.3 s), for a total improvement by a factor of almost 12.
Thanks to computational, algorithmic, and programming innovations, our tool can handle unprecedentedly large atomic systems (up to 84,480 atoms) and achieve a sustained Eflop/s performance in FP64. In particular, it accounts for dissipative mechanisms that are expected to drastically impact the behavior of nano-transistors, e.g., NRFETs. Although the current implementation includes electron-electron interactions only, other types of scattering, such as electron-phonon or electron-photon, can be readily integrated, making our package the reference in the technology computer-aided design (TCAD) of ultra-scaled components.

Nevertheless, some limitations persist.
Handling realistically-sized devices consisting of more than 25,000 atoms requires the use of distributed-memory algorithms to solve for the Green's functions and screened Coulomb interactions.
Consequently, the size of the energy grid cannot be as large as in the case of structure, e.g., NW-1 and NW-2, where multiple energies fit into a single GPU.
For example, even at the full scale of Frontier, we could ``only'' simulate $\sim$19-38$\times10^3$ energies, which is at the lower bound of the necessary resolution. A number closer to 100,000 would be better.

We believe there are further memory-related innovations to be exploited that can bring us closer towards the goal of combining both realistically-sized devices and energy grids on the current generation of supercomputers.
In this work, we focused on keeping data on the GPU memory to minimize data movement and maximize the compute potential. Our memory-handling scheme can, therefore, be further improved by making better use of the host memory.
Also, the data that has to be stored and communicated to the energy convolutions can potentially be reduced by appropriate compression or lower-precision schemes that retain the required computational accuracy.
Overall, we are planning to explore these optimization opportunities and bring our ``nano-TCAD'' tool \mbox{QuaTrEx} to the next level of size and accuracy.

\section*{Acknowledgment}
This work was supported by the Swiss National Science Foundation (SNSF) under grant $\mathrm{n^\circ}$ 209358 (QuaTrEx) and grant $\mathrm{n^\circ}$~205602 (NCCR MARVEL), and by the Platform for Advanced Scientific Computing in Switzerland (BoostQT). We acknowledge support from CSCS (projects c33, g34, g186, lp16, lp82). This research used resources of the Oak Ridge Leadership Computing Facility at the Oak Ridge National Laboratory, which is supported by the Office of Science of the U.S. Department of Energy under Contract No. DE-AC05-00OR22725 (project NEL107). We especially thank Maria Grazia Giuffreda, Ver\'{o}nica G. Melesse Vergara, and Rocco Meli.

\let\oldthebibliography\thebibliography
\renewcommand{\thebibliography}[1]{%
  \oldthebibliography{#1}%
  \footnotesize
}

\bibliographystyle{ACM-Reference-Format}
\bibliography{bibliography}


\begin{thebibliography}{52}


\ifx \showCODEN    \undefined \def \showCODEN     #1{\unskip}     \fi
\ifx \showISBNx    \undefined \def \showISBNx     #1{\unskip}     \fi
\ifx \showISBNxiii \undefined \def \showISBNxiii  #1{\unskip}     \fi
\ifx \showISSN     \undefined \def \showISSN      #1{\unskip}     \fi
\ifx \showLCCN     \undefined \def \showLCCN      #1{\unskip}     \fi
\ifx \shownote     \undefined \def \shownote      #1{#1}          \fi
\ifx \showarticletitle \undefined \def \showarticletitle #1{#1}   \fi
\ifx \showURL      \undefined \def \showURL       {\relax}        \fi
\providecommand\bibfield[2]{#2}
\providecommand\bibinfo[2]{#2}
\providecommand\natexlab[1]{#1}
\providecommand\showeprint[2][]{arXiv:#2}

\bibitem[Agrawal et~al\mbox{.}(2024)]%
        {intelnr}
\bibfield{author}{\bibinfo{person}{A. Agrawal}  {et~al\mbox{.}}} \bibinfo{year}{2024}\natexlab{}.
\newblock \showarticletitle{Silicon RibbonFET CMOS at 6nm Gate Length}. In \bibinfo{booktitle}{\emph{2024 IEEE International Electron Devices Meeting (IEDM)}}.


\bibitem[Artacho et~al\mbox{.}(2008)]%
        {siesta}
\bibfield{author}{\bibinfo{person}{Emilio Artacho}  {et~al\mbox{.}}} \bibinfo{year}{2008}\natexlab{}.
\newblock \showarticletitle{The SIESTA method; developments and applicability}.
\newblock \bibinfo{journal}{\emph{Journal of Physics: Condensed Matter}} (\bibinfo{date}{Jan.} \bibinfo{year}{2008}).


\bibitem[Atchley et~al\mbox{.}(2023)]%
        {Atchley2023_frontier}
\bibfield{author}{\bibinfo{person}{Scott Atchley}  {et~al\mbox{.}}} \bibinfo{year}{2023}\natexlab{}.
\newblock \showarticletitle{Frontier: Exploring Exascale}. In \bibinfo{booktitle}{\emph{Proceedings of the International Conference for High Performance Computing, Networking, Storage and Analysis}} \emph{(\bibinfo{series}{SC ’23})}. \bibinfo{publisher}{ACM}.


\bibitem[Ben et~al\mbox{.}(2020)]%
        {berkeleygw}
\bibfield{author}{\bibinfo{person}{Mauro~Del Ben}  {et~al\mbox{.}}} \bibinfo{year}{2020}\natexlab{}.
\newblock \showarticletitle{Accelerating {{Large-Scale Excited-State GW Calculations}} on {{Leadership HPC Systems}}}. In \bibinfo{booktitle}{\emph{Proc. of {{SC20}}: {{International Conference}} for {{High Performance Computing}}, {{Networking}}, {{Storage}} and {{Analysis}}}}.


\bibitem[Beyn(2012)]%
        {beynIntegralMethodSolving2012}
\bibfield{author}{\bibinfo{person}{Wolf-J{\"u}rgen Beyn}.} \bibinfo{year}{2012}\natexlab{}.
\newblock \showarticletitle{An Integral Method for Solving Nonlinear Eigenvalue Problems}.
\newblock \bibinfo{journal}{\emph{Linear Algebra Appl.}} (\bibinfo{date}{May} \bibinfo{year}{2012}).


\bibitem[Bowen et~al\mbox{.}(1995)]%
        {Bowen1995}
\bibfield{author}{\bibinfo{person}{R.~Chris Bowen}  {et~al\mbox{.}}} \bibinfo{year}{1995}\natexlab{}.
\newblock \showarticletitle{Transmission resonances and zeros in multiband models}.
\newblock \bibinfo{journal}{\emph{Physical Review B}} (\bibinfo{date}{July} \bibinfo{year}{1995}).


\bibitem[Brandbyge et~al\mbox{.}(2002)]%
        {stokbro}
\bibfield{author}{\bibinfo{person}{Mads Brandbyge}  {et~al\mbox{.}}} \bibinfo{year}{2002}\natexlab{}.
\newblock \showarticletitle{Density-functional method for nonequilibrium electron transport}.
\newblock \bibinfo{journal}{\emph{Phys. Rev. B}} (\bibinfo{date}{March} \bibinfo{year}{2002}).


\bibitem[Br{\"u}ck et~al\mbox{.}(2017)]%
        {bruckEfficientAlgorithmsLargescale2017}
\bibfield{author}{\bibinfo{person}{Sascha Br{\"u}ck}  {et~al\mbox{.}}} \bibinfo{year}{2017}\natexlab{}.
\newblock \showarticletitle{Efficient Algorithms for Large-Scale Quantum Transport Calculations}.
\newblock \bibinfo{journal}{\emph{The Journal of Chemical Physics}} (\bibinfo{date}{Aug.} \bibinfo{year}{2017}).


\bibitem[Calderara et~al\mbox{.}(2015)]%
        {calderara}
\bibfield{author}{\bibinfo{person}{Mauro Calderara}  {et~al\mbox{.}}} \bibinfo{year}{2015}\natexlab{}.
\newblock \showarticletitle{Pushing back the limit of ab-initio quantum transport simulations on hybrid supercomputers}. In \bibinfo{booktitle}{\emph{Proceedings of the International Conference for High Performance Computing, Networking, Storage and Analysis}} (Austin, Texas) \emph{(\bibinfo{series}{SC '15})}. \bibinfo{publisher}{Association for Computing Machinery}.


\bibitem[Callen and Welton(1951)]%
        {fluctuation}
\bibfield{author}{\bibinfo{person}{Herbert~B. Callen} {and} \bibinfo{person}{Theodore~A. Welton}.} \bibinfo{year}{1951}\natexlab{}.
\newblock \showarticletitle{Irreversibility and Generalized Noise}.
\newblock \bibinfo{journal}{\emph{Phys. Rev.}} (\bibinfo{date}{July} \bibinfo{year}{1951}).


\bibitem[Cauley et~al\mbox{.}(2011)]%
        {pdiv}
\bibfield{author}{\bibinfo{person}{Stephen Cauley}  {et~al\mbox{.}}} \bibinfo{year}{2011}\natexlab{}.
\newblock \showarticletitle{Distributed non-equilibrium {Green}’s function algorithms for the simulation of nanoelectronic devices with scattering}.
\newblock \bibinfo{journal}{\emph{Journal of Applied Physics}} (\bibinfo{date}{Aug.} \bibinfo{year}{2011}).


\bibitem[Chen and Pasquarello(2017)]%
        {Chen2017}
\bibfield{author}{\bibinfo{person}{Wei Chen} {and} \bibinfo{person}{Alfredo Pasquarello}.} \bibinfo{year}{2017}\natexlab{}.
\newblock \showarticletitle{Accuracy of GW for calculating defect energy levels in solids}.
\newblock \bibinfo{journal}{\emph{Physical Review B}} (\bibinfo{date}{July} \bibinfo{year}{2017}).


\bibitem[Dalcin and Fang(2021)]%
        {dalcinMpi4pyStatusUpdate2021}
\bibfield{author}{\bibinfo{person}{Lisandro Dalcin} {and} \bibinfo{person}{Yao-Lung~L. Fang}.} \bibinfo{year}{2021}\natexlab{}.
\newblock \showarticletitle{Mpi4py: {{Status Update After}} 12 {{Years}} of {{Development}}}.
\newblock \bibinfo{journal}{\emph{Computing in Science \& Engineering}} (\bibinfo{date}{July} \bibinfo{year}{2021}).


\bibitem[Deuschle et~al\mbox{.}(2025)]%
        {deuschle2024_arXiv}
\bibfield{author}{\bibinfo{person}{Leonard Deuschle}  {et~al\mbox{.}}} \bibinfo{year}{2025}\natexlab{}.
\newblock \showarticletitle{Electron-electron interactions in device simulation via nonequilibrium Green's functions and the GW approximation}.
\newblock \bibinfo{journal}{\emph{Phys. Rev. B}} (\bibinfo{date}{May} \bibinfo{year}{2025}).


\bibitem[Deuschle et~al\mbox{.}(2024)]%
        {deuschle2024_sc}
\bibfield{author}{\bibinfo{person}{Leonard Deuschle}  {et~al\mbox{.}}} \bibinfo{year}{2024}\natexlab{}.
\newblock \showarticletitle{Towards Exascale Simulations of Nanoelectronic Devices in the GW Approximation}. In \bibinfo{booktitle}{\emph{Proceedings of the International Conference for High Performance Computing, Networking, Storage, and Analysis}} (Atlanta, GA, USA) \emph{(\bibinfo{series}{SC '24})}. \bibinfo{publisher}{IEEE Press}.


\bibitem[docs.lumi supercomputer.eu(2025)]%
        {lumi}
\bibfield{author}{\bibinfo{person}{docs.lumi supercomputer.eu}.} \bibinfo{year}{2025}\natexlab{}.
\newblock \bibinfo{title}{GPU nodes - LUMI-G}.


\bibitem[Efron(1992)]%
        {bootstrapping}
\bibfield{author}{\bibinfo{person}{Bradley Efron}.} \bibinfo{year}{1992}\natexlab{}.
\newblock \showarticletitle{Bootstrap {{Methods}}: {{Another Look}} at the {{Jackknife}}}.
\newblock In \bibinfo{booktitle}{\emph{Breakthroughs in {{Statistics}}: {{Methodology}} and {{Distribution}}}}, \bibfield{editor}{\bibinfo{person}{Samuel Kotz} {and} \bibinfo{person}{Norman~L. Johnson}} (Eds.). \bibinfo{publisher}{Springer}.


\bibitem[Fischetti(2001)]%
        {Fischetti2001}
\bibfield{author}{\bibinfo{person}{M.~V. Fischetti}.} \bibinfo{year}{2001}\natexlab{}.
\newblock \showarticletitle{Long-range Coulomb interactions in small Si devices. Part II. Effective electron mobility in thin-oxide structures}.
\newblock \bibinfo{journal}{\emph{Journal of Applied Physics}} (\bibinfo{date}{Jan.} \bibinfo{year}{2001}).


\bibitem[Fusco et~al\mbox{.}(2024)]%
        {hoefler_alps}
\bibfield{author}{\bibinfo{person}{Luigi Fusco}  {et~al\mbox{.}}} \bibinfo{year}{2024}\natexlab{}.
\newblock \bibinfo{title}{Understanding Data Movement in Tightly Coupled Heterogeneous Systems: A Case Study with the Grace Hopper Superchip}.


\bibitem[Gao et~al\mbox{.}(2024)]%
        {nanogw}
\bibfield{author}{\bibinfo{person}{Weiwei Gao}  {et~al\mbox{.}}} \bibinfo{year}{2024}\natexlab{}.
\newblock \showarticletitle{Efficient Full-Frequency GW Calculations Using a Lanczos Method}.
\newblock \bibinfo{journal}{\emph{Phys. Rev. Lett.}} (\bibinfo{date}{March} \bibinfo{year}{2024}).


\bibitem[Giannozzi et~al\mbox{.}(2017)]%
        {qe}
\bibfield{author}{\bibinfo{person}{P Giannozzi}  {et~al\mbox{.}}} \bibinfo{year}{2017}\natexlab{}.
\newblock \showarticletitle{Advanced capabilities for materials modelling with QUANTUM ESPRESSO}.
\newblock \bibinfo{journal}{\emph{Journal of Physics: Condensed Matter}} (\bibinfo{date}{Oct.} \bibinfo{year}{2017}).


\bibitem[Gonze et~al\mbox{.}(2020)]%
        {abinit}
\bibfield{author}{\bibinfo{person}{Xavier Gonze}  {et~al\mbox{.}}} \bibinfo{year}{2020}\natexlab{}.
\newblock \showarticletitle{The Abinit project: Impact, environment and recent developments}.
\newblock \bibinfo{journal}{\emph{Comput. Phys. Commun.}} (\bibinfo{date}{March} \bibinfo{year}{2020}).


\bibitem[Harris et~al\mbox{.}(2020)]%
        {harris2020array}
\bibfield{author}{\bibinfo{person}{Charles~R. Harris}  {et~al\mbox{.}}} \bibinfo{year}{2020}\natexlab{}.
\newblock \showarticletitle{Array programming with {NumPy}}.
\newblock \bibinfo{journal}{\emph{Nature}} (\bibinfo{date}{Sept.} \bibinfo{year}{2020}).


\bibitem[Hedin(1965)]%
        {Hedin1965}
\bibfield{author}{\bibinfo{person}{Lars Hedin}.} \bibinfo{year}{1965}\natexlab{}.
\newblock \showarticletitle{New Method for Calculating the One-Particle Green's Function with Application to the Electron-Gas Problem}.
\newblock \bibinfo{journal}{\emph{Physical Review}} (\bibinfo{date}{Sept.} \bibinfo{year}{1965}).


\bibitem[Hybertsen and Louie(1986)]%
        {Hybertsen1986}
\bibfield{author}{\bibinfo{person}{Mark~S. Hybertsen} {and} \bibinfo{person}{Steven~G. Louie}.} \bibinfo{year}{1986}\natexlab{}.
\newblock \showarticletitle{Electron correlation in semiconductors and insulators: Band gaps and quasiparticle energies}.
\newblock \bibinfo{journal}{\emph{Physical Review B}} (\bibinfo{date}{Oct.} \bibinfo{year}{1986}).


\bibitem[Kitagawa(1977)]%
        {Kitagawa1977}
\bibfield{author}{\bibinfo{person}{Genshiro Kitagawa}.} \bibinfo{year}{1977}\natexlab{}.
\newblock \showarticletitle{An algorithm for solving the matrix equation $X = FXF^T + S$}.
\newblock \bibinfo{journal}{\emph{Internat. J. Control}} (\bibinfo{date}{Jan.} \bibinfo{year}{1977}).


\bibitem[Kohn and Sham(1965)]%
        {kohnsham}
\bibfield{author}{\bibinfo{person}{W. Kohn} {and} \bibinfo{person}{L.~J. Sham}.} \bibinfo{year}{1965}\natexlab{}.
\newblock \showarticletitle{Self-Consistent Equations Including Exchange and Correlation Effects}.
\newblock \bibinfo{journal}{\emph{Phys. Rev.}} (\bibinfo{date}{Nov.} \bibinfo{year}{1965}).


\bibitem[Kresse and Hafner(1993)]%
        {vasp}
\bibfield{author}{\bibinfo{person}{G. Kresse} {and} \bibinfo{person}{J. Hafner}.} \bibinfo{year}{1993}\natexlab{}.
\newblock \showarticletitle{Ab initio molecular dynamics for liquid metals}.
\newblock \bibinfo{journal}{\emph{Phys. Rev. B}} (\bibinfo{date}{Jan.} \bibinfo{year}{1993}).


\bibitem[Kutepov(2020)]%
        {scgw}
\bibfield{author}{\bibinfo{person}{A.~L. Kutepov}.} \bibinfo{year}{2020}\natexlab{}.
\newblock \showarticletitle{Self-Consistent {{GW}} Method: {{O}}({{N}}) Algorithm for Polarizability and Self Energy}.
\newblock \bibinfo{journal}{\emph{Comput. Phys. Commun.}} (\bibinfo{date}{Dec.} \bibinfo{year}{2020}).


\bibitem[Kühne et~al\mbox{.}(2020)]%
        {cp2k}
\bibfield{author}{\bibinfo{person}{Thomas~D. Kühne}  {et~al\mbox{.}}} \bibinfo{year}{2020}\natexlab{}.
\newblock \showarticletitle{CP2K: An electronic structure and molecular dynamics software package - Quickstep: Efficient and accurate electronic structure calculations}.
\newblock \bibinfo{journal}{\emph{The Journal of Chemical Physics}} (\bibinfo{date}{May} \bibinfo{year}{2020}).


\bibitem[Lake et~al\mbox{.}(1997)]%
        {rgf}
\bibfield{author}{\bibinfo{person}{Roger Lake}  {et~al\mbox{.}}} \bibinfo{year}{1997}\natexlab{}.
\newblock \showarticletitle{Single and multiband modeling of quantum electron transport through layered semiconductor devices}.
\newblock \bibinfo{journal}{\emph{Journal of Applied Physics}} (\bibinfo{date}{June} \bibinfo{year}{1997}).


\bibitem[Lam et~al\mbox{.}(2015)]%
        {lam2015numba}
\bibfield{author}{\bibinfo{person}{Siu~Kwan Lam}, \bibinfo{person}{Antoine Pitrou}  {and} \bibinfo{person}{Stanley Seibert}.} \bibinfo{year}{2015}\natexlab{}.
\newblock \showarticletitle{Numba: A llvm-based python jit compiler}. In \bibinfo{booktitle}{\emph{Proceedings of the Second Workshop on the LLVM Compiler Infrastructure in HPC}}.


\bibitem[Liu et~al\mbox{.}(2016)]%
        {vaspgw}
\bibfield{author}{\bibinfo{person}{Peitao Liu}  {et~al\mbox{.}}} \bibinfo{year}{2016}\natexlab{}.
\newblock \showarticletitle{Cubic Scaling {$GW$}: {{Towards}} Fast Quasiparticle Calculations}.
\newblock \bibinfo{journal}{\emph{Phys. Rev. B}} (\bibinfo{date}{Oct.} \bibinfo{year}{2016}).


\bibitem[Luisier et~al\mbox{.}(2006)]%
        {luisierAtomisticSimulationNanowires2006}
\bibfield{author}{\bibinfo{person}{Mathieu Luisier}  {et~al\mbox{.}}} \bibinfo{year}{2006}\natexlab{}.
\newblock \showarticletitle{Atomistic Simulation of Nanowires in the $sp^3d^5s^*$ Tight-Binding Formalism: {{From}} Boundary Conditions to Strain Calculations}.
\newblock \bibinfo{journal}{\emph{Physical Review B}} (\bibinfo{date}{Nov.} \bibinfo{year}{2006}).


\bibitem[Maillou et~al\mbox{.}(2025)]%
        {maillouSerinvScalableLibrary2025}
\bibfield{author}{\bibinfo{person}{Vincent Maillou}  {et~al\mbox{.}}} \bibinfo{year}{2025}\natexlab{}.
\newblock \bibinfo{title}{Serinv: {{A Scalable Library}} for the {{Selected Inversion}} of {{Block-Tridiagonal}} with {{Arrowhead Matrices}}}.


\bibitem[Marzari and Vanderbilt(1997)]%
        {mlwf}
\bibfield{author}{\bibinfo{person}{Nicola Marzari} {and} \bibinfo{person}{David Vanderbilt}.} \bibinfo{year}{1997}\natexlab{}.
\newblock \showarticletitle{Maximally localized generalized Wannier functions for composite energy bands}.
\newblock \bibinfo{journal}{\emph{Phys. Rev. B}} (\bibinfo{date}{Nov.} \bibinfo{year}{1997}).


\bibitem[Mortensen et~al\mbox{.}(2024)]%
        {GPAW}
\bibfield{author}{\bibinfo{person}{Jens~Jørgen Mortensen}  {et~al\mbox{.}}} \bibinfo{year}{2024}\natexlab{}.
\newblock \showarticletitle{GPAW: An open Python package for electronic structure calculations}.
\newblock \bibinfo{journal}{\emph{The Journal of Chemical Physics}} (\bibinfo{date}{March} \bibinfo{year}{2024}).


\bibitem[Okuta et~al\mbox{.}(2017)]%
        {cupy_learningsys2017}
\bibfield{author}{\bibinfo{person}{Ryosuke Okuta}  {et~al\mbox{.}}} \bibinfo{year}{2017}\natexlab{}.
\newblock \showarticletitle{CuPy: A NumPy-Compatible Library for NVIDIA GPU Calculations}. In \bibinfo{booktitle}{\emph{Proceedings of Workshop on Machine Learning Systems (LearningSys) in The Thirty-first Annual Conference on Neural Information Processing Systems (NIPS)}}.


\bibitem[Onida et~al\mbox{.}(2002)]%
        {Onida2002}
\bibfield{author}{\bibinfo{person}{G. Onida}, \bibinfo{person}{L. Reining}  {and} \bibinfo{person}{A. Rubio}.} \bibinfo{year}{2002}\natexlab{}.
\newblock \showarticletitle{Many-body perturbation theory approaches to electronic excitations: density functional versus Green function methods}.
\newblock \bibinfo{journal}{\emph{Reviews of Modern Physics}} (\bibinfo{date}{June} \bibinfo{year}{2002}).


\bibitem[Perdew et~al\mbox{.}(2009)]%
        {Perdew2009}
\bibfield{author}{\bibinfo{person}{John~P. Perdew}  {et~al\mbox{.}}} \bibinfo{year}{2009}\natexlab{}.
\newblock \showarticletitle{Some Fundamental Issues in Ground-State Density Functional Theory: A Guide for the Perplexed}.
\newblock \bibinfo{journal}{\emph{Journal of Chemical Theory and Computation}} (\bibinfo{date}{April} \bibinfo{year}{2009}).


\bibitem[Petersen et~al\mbox{.}(2009)]%
        {psr}
\bibfield{author}{\bibinfo{person}{Dan~Erik Petersen}  {et~al\mbox{.}}} \bibinfo{year}{2009}\natexlab{}.
\newblock \showarticletitle{A hybrid method for the parallel computation of {Green}’s functions}.
\newblock \bibinfo{journal}{\emph{J. Comput. Phys.}} (\bibinfo{date}{Aug.} \bibinfo{year}{2009}).


\bibitem[Pizzi et~al\mbox{.}(2020)]%
        {Wannier90}
\bibfield{author}{\bibinfo{person}{Giovanni Pizzi}  {et~al\mbox{.}}} \bibinfo{year}{2020}\natexlab{}.
\newblock \showarticletitle{Wannier90 as a Community Code: New Features and Applications}.
\newblock \bibinfo{journal}{\emph{Journal of Physics: Condensed Matter}} (\bibinfo{date}{April} \bibinfo{year}{2020}).


\bibitem[Poloni(2020)]%
        {Poloni2020}
\bibfield{author}{\bibinfo{person}{Federico Poloni}.} \bibinfo{year}{2020}\natexlab{}.
\newblock \showarticletitle{Iterative and doubling algorithms for Riccati‐type matrix equations: A comparative introduction}.
\newblock \bibinfo{journal}{\emph{GAMM-Mitteilungen}} (\bibinfo{date}{Oct.} \bibinfo{year}{2020}).


\bibitem[Qiu et~al\mbox{.}(2013)]%
        {Qiu2013}
\bibfield{author}{\bibinfo{person}{Diana~Y. Qiu}, \bibinfo{person}{Felipe~H. da Jornada}  {and} \bibinfo{person}{Steven~G. Louie}.} \bibinfo{year}{2013}\natexlab{}.
\newblock \showarticletitle{Optical Spectrum of MoS$_2$: Many-Body Effects and Diversity of Exciton States}.
\newblock \bibinfo{journal}{\emph{Physical Review Letters}} (\bibinfo{date}{Nov.} \bibinfo{year}{2013}).


\bibitem[Sancho et~al\mbox{.}(1985)]%
        {sancho}
\bibfield{author}{\bibinfo{person}{M~P~Lopez Sancho}  {et~al\mbox{.}}} \bibinfo{year}{1985}\natexlab{}.
\newblock \showarticletitle{Highly Convergent Schemes for the Calculation of Bulk and Surface {{Green}} Functions}.
\newblock \bibinfo{journal}{\emph{Journal of Physics F: Metal Physics}} (\bibinfo{date}{April} \bibinfo{year}{1985}).


\bibitem[Thygesen and Rubio(2007)]%
        {thygesen2007}
\bibfield{author}{\bibinfo{person}{Kristian~S. Thygesen} {and} \bibinfo{person}{Angel Rubio}.} \bibinfo{year}{2007}\natexlab{}.
\newblock \showarticletitle{Nonequilibrium GW approach to quantum transport in nano-scale contacts}.
\newblock \bibinfo{journal}{\emph{The Journal of Chemical Physics}} (\bibinfo{date}{March} \bibinfo{year}{2007}).


\bibitem[top500.org(2024)]%
        {top500}
\bibfield{author}{\bibinfo{person}{top500.org}.} \bibinfo{year}{2024}\natexlab{}.
\newblock \bibinfo{title}{TOP500 November 2024}.


\bibitem[Virtanen et~al\mbox{.}(2020)]%
        {2020SciPy-NMeth}
\bibfield{author}{\bibinfo{person}{Pauli Virtanen}  {et~al\mbox{.}}} \bibinfo{year}{2020}\natexlab{}.
\newblock \showarticletitle{{{SciPy} 1.0: Fundamental Algorithms for Scientific Computing in Python}}.
\newblock \bibinfo{journal}{\emph{Nature Methods}} (\bibinfo{date}{Feb.} \bibinfo{year}{2020}).


\bibitem[Wilhelm et~al\mbox{.}(2018)]%
        {cp2kgw}
\bibfield{author}{\bibinfo{person}{Jan Wilhelm}  {et~al\mbox{.}}} \bibinfo{year}{2018}\natexlab{}.
\newblock \showarticletitle{Toward {{GW Calculations}} on {{Thousands}} of {{Atoms}}}.
\newblock \bibinfo{journal}{\emph{J. Phys. Chem. Lett.}} (\bibinfo{date}{Jan.} \bibinfo{year}{2018}).


\bibitem[Wu et~al\mbox{.}(2024)]%
        {gwsc24}
\bibfield{author}{\bibinfo{person}{Wentiao Wu}  {et~al\mbox{.}}} \bibinfo{year}{2024}\natexlab{}.
\newblock \showarticletitle{Enabling 13K-Atom Excited-State GW Calculations via Low-Rank Approximations and HPC on the New Sunway Supercomputer}. In \bibinfo{booktitle}{\emph{SC24: International Conference for High Performance Computing, Networking, Storage and Analysis}}.


\bibitem[Yu and Govoni(2022)]%
        {westgw}
\bibfield{author}{\bibinfo{person}{Victor Wen-zhe Yu} {and} \bibinfo{person}{Marco Govoni}.} \bibinfo{year}{2022}\natexlab{}.
\newblock \showarticletitle{GPU Acceleration of Large-Scale Full-Frequency GW Calculations}.
\newblock \bibinfo{journal}{\emph{Journal of Chemical Theory and Computation}} (\bibinfo{date}{Aug.} \bibinfo{year}{2022}).


\bibitem[Ziogas et~al\mbox{.}(2019)]%
        {ziogas}
\bibfield{author}{\bibinfo{person}{Alexandros~Nikolaos Ziogas}  {et~al\mbox{.}}} \bibinfo{year}{2019}\natexlab{}.
\newblock \showarticletitle{A data-centric approach to extreme-scale ab initio dissipative quantum transport simulations}. In \bibinfo{booktitle}{\emph{Proceedings of the International Conference for High Performance Computing, Networking, Storage and Analysis}} (Denver, Colorado) \emph{(\bibinfo{series}{SC '19})}. \bibinfo{publisher}{Association for Computing Machinery}.


\end{thebibliography}

\end{document}